\documentclass[english,prd,twocolumn,superscriptaddress,showpacs,preprintnumbers,nofootinbib]{revtex4-1}

\usepackage{graphicx}
\usepackage{dcolumn}
\usepackage{bm}
\usepackage{amsmath}
\usepackage{amssymb}
\usepackage{pifont}
\usepackage{tabularx} 
\usepackage{etoolbox}
\usepackage{booktabs} 
\usepackage{float}
\usepackage[dvipsnames, svgnames]{xcolor}
\usepackage{hyperref}
\usepackage{xcolor}

\hypersetup{
    colorlinks=true,
    linkcolor=blue,
    citecolor=blue,
    urlcolor=blue
}

\begin{document}

\title{Impact of reactor antineutrinos on the neutrino floor in low-mass WIMP-like dark matter searches}





\author{S.~Das}
\email{sudipta.das@niser.ac.in}
\author{V.~K.~S.~Kashyap}
\author{B.~Mohanty}
\affiliation{National Institute of Science Education and Research, an OCC of Homi Bhabha National Institute, Jatni 752050, India}

\date{\today}

\begin{abstract}

The sensitivity of conventional direct dark matter searches for weakly interacting massive particles (WIMPs) is ultimately limited by coherent elastic neutrino–nucleus scattering (CE$\nu$NS), which produces nuclear recoils indistinguishable from WIMP signals and defines the so-called \emph{neutrino floor}.
While the effects of solar neutrinos, geoneutrinos, diffuse supernova neutrinos, and atmospheric neutrinos have been extensively studied in this context, the contribution from reactor antineutrinos has received comparatively little attention.
We present the first systematic evaluation of how reactor antineutrino fluxes, modeled as a function of reactor–detector distance, modify the neutrino floor for low-mass WIMP searches using SuperCDMS-like high-voltage germanium detectors.
Both discovery-limit and opacity-based formulations of the neutrino floor are examined under consistent assumptions.
We find that proximity to gigawatt-scale reactors within $\sim$10~km can raise the neutrino floor by up to a few orders of magnitude, significantly reducing the sensitivity to sub-10~GeV/$c^{2}$ dark matter.
Beyond $\sim$100~km, the reactor contribution becomes negligible. These conclusions hold for both definitions of the neutrino floor and remain stable under reasonable variations in detector quenching, site-dependent geoneutrino flux, and reactor antineutrino flux uncertainties, emphasizing reactor proximity as a critical factor in site selection for future low-threshold dark matter experiments.

\end{abstract}

\maketitle

\section{Introduction}

The existence of dark matter (DM) is supported by a wide range of astrophysical and cosmological observations~\cite{first_proposed_DM:1933gu,rotation_curve_1970,bullet_cluster_2006,Planck_2015_CMB}. 
However, its particle nature remains unknown. Weakly Interacting Massive Particles (WIMPs) are theoretically well-motivated and represent one of the leading dark matter candidates~\cite{WIMP_1_1984,WIMP_2_1995_supersymmetric,Relic_WIMP_abundances_2012}. 
Direct detection experiments search for nuclear recoils produced by the elastic scattering of WIMPs off target nuclei. 
Since both the WIMP mass ($M_\chi$) and the WIMP-nucleon scattering cross section ($\sigma_{\chi-n}$) are unknown, these experiments explore a wide parameter space in $M_\chi$ and $\sigma_{\chi-n}$~\cite{CDMSliteR3_WIMP_PLR_2019,CRESST_2024_WIMP,LZ_2023_WIMP,PandaX_2025_WIMP,XENONnT_2025_WIMP}.

Next-generation direct detection experiments are rapidly approaching sensitivities to very small cross sections~\cite{SuperCDMS_SNOALB_projected_SNOWMASS_2021,CRESST_2025_projected_sensitivity,xenon2024xenonnt_status}. 
At these levels, an irreducible background from coherent elastic neutrino–nucleus scattering (CE$\nu$NS)~\cite{freedman} becomes a limiting factor. 
In CE$\nu$NS, neutrinos scatter coherently off entire nuclei, producing nuclear recoils that are experimentally indistinguishable from those expected from WIMP interactions. 
The smallest WIMP cross section discoverable under this background corresponds to the point at which a WIMP signal would produce a statistically significant excess (typically $3\sigma$) over the CE$\nu$NS background. 
Below this level, the dark matter signal becomes effectively obscured, defining the so-called \emph{neutrino floor}~\cite{Billard:2013qya,Ruppin_followup:2014bra,Overcome_neutrino_floor:2020lva}.

The neutrino floor is not a strict boundary but rather a sensitivity threshold that depends on several factors, including the theoretical understanding of CE$\nu$NS cross sections, the neutrino fluxes from various sources, and the total exposure of experiments. To emphasize its diffuse and site-dependent nature, the term \emph{neutrino fog} has also been introduced~\cite{OHare:2021utq,CJPL_neutrino_floor_Fan:2025sde}. Although recent CE$\nu$NS measurements are consistent with Standard Model predictions~\cite{COHERENT_CEvNS,COUNSplus_CEvNS,COHERENT:Ge_2025}, the dominant uncertainty in determining the neutrino floor arises from systematic uncertainties in the neutrino fluxes, which can reach up to 50\%.

Neutrino-induced backgrounds arise from both natural and man-made sources. Natural backgrounds include solar neutrinos, geoneutrinos, the diffuse supernova neutrino background (DSNB), and atmospheric neutrinos, while man-made contributions mainly originate from nuclear reactor operations. The flux of reactor electron antineutrinos can be substantial, particularly for experiments located near reactor facilities. Although solar and atmospheric neutrino backgrounds are well characterized in the context of the neutrino floor, the role of reactor antineutrinos has traditionally received less attention~\cite{Billard:2013qya,OHare:2021utq}. For searches targeting low-mass WIMPs in the sub-GeV/$c^{2}$ to GeV/$c^{2}$ mass range, CE$\nu$NS induced by reactor antineutrinos can significantly affect sensitivity, depending on the experiment’s distance from the nearest reactor.

Germanium-based cryogenic detectors operated at high bias voltages (HV) and achieving ultra-low energy thresholds, such as those employed by SuperCDMS~\cite{CDMSliteR3_WIMP_PLR_2019}, are particularly well suited for probing WIMPs in this mass regime. Therefore, our study focuses on determining the neutrino floor for germanium-based detectors. We systematically quantify the impact of reactor antineutrino fluxes on the neutrino floor. Using detailed spectral models of a GW$_{\text{th}}$ reactor configuration and accounting for the detector quenching response, we evaluate how CE$\nu$NS-induced backgrounds vary with distance from the reactor and how they influence WIMP discovery limits using two widely adopted definitions of the neutrino floor: the discovery-based and opacity-based approaches.

The pioneering works by Billard \textit{et al.}~\cite{Billard:2013qya} and Ruppin \textit{et al.}~\cite{Ruppin_followup:2014bra} established the discovery-limit-based definition of the neutrino floor for xenon, silicon, and germanium targets, without accounting for reactor or geoneutrino fluxes. Subsequently, O’Hare~\cite{OHare:2021utq} introduced the opacity (or “neutrino fog”) framework, which included reactor backgrounds only for the Gran Sasso site, but without exploring its distance dependence. More recently, Fan \textit{et al.}~\cite{CJPL_neutrino_floor_Fan:2025sde} evaluated the opacity-based neutrino floor at the China Jinping Underground Laboratory (CJPL), without considering distance-dependent reactor $\bar{\nu}_e$ effects, while D. Aristizabal Sierra \textit{et al.}~\cite{reactor_anti_neutrino_contribution_PRD_2024} studied site-specific reactor antineutrino fluxes across different underground laboratories but did not assess their impact on the neutrino floor. 

The present work builds upon these foundational studies by providing the quantitative assessment of how reactor $\bar{\nu}_e$ fluxes—explicitly modeled as a function of reactor–detector distance—affect both discovery-limit-based and opacity-based neutrino floors for germanium detectors. This study thus establishes, for the first time, the explicit distance dependence of the neutrino floor and highlights the importance of site-specific reactor antineutrino modeling in future dark matter sensitivity projections.

Our results show that proximity to a gigawatt-scale reactor within ten kilometers can elevate the neutrino floor by up to a few orders of magnitude across the low-mass WIMP region (in the few~GeV/$c^2$ range), compared to baseline ``no-reactor'' studies~\cite{Billard:2013qya,Billard:APPEC_report,Ruppin_followup:2014bra} or those assuming reactor–detector separations of $\mathcal{O}(100)$~km~\cite{OHare:2021utq}. This substantial enhancement has important implications for the design and siting of next-generation direct detection experiments targeting sub-10~GeV/$c^2$ dark matter. 

The paper is organized as follows. 
Section~\ref{sec:WIMP_signal_modeling} presents the modeling of the WIMP signal. 
Section~\ref{sec:CEvNS_background_modeling} describes the modeling of CE$\nu$NS-induced backgrounds. 
In Section~\ref{sec:neutrino_floor_definition}, we examine the impact of reactor antineutrinos on the neutrino floor using two widely adopted definitions. 
Section~\ref{Impact in next generation experiment} discusses the implications of these effects for future dark matter experiments. 
Finally, Section~\ref{Conclusion} summarizes the main findings and conclusions.

\section{WIMP Signal Rate Modeling \label{sec:WIMP_signal_modeling}}

\subsection{Differential Event Rate}

Direct detection experiments primarily aim to measure the tiny nuclear recoil energy $E_R$ imparted to a target nucleus through the elastic scattering of a WIMP.  
The differential event rate per unit detector mass can be expressed as~\cite{WIMP_rate_Lewin}
\begin{equation}
\frac{dR}{dE_R} = 
\frac{N_A \, \rho_0}{A} 
\frac{\sigma_N}{2 M_\chi \, \mu_N^2} 
F^2(E_R) \, \eta(v_{\min}),
\label{eq:dRdE_sigmaN_journal}
\end{equation}
where $N_A$ is Avogadro’s number, $\rho_0$ is the local dark matter density, $A$ is the atomic mass number of the target nucleus, and $M_\chi$ is the WIMP mass.  
The WIMP--nucleus reduced mass is $\mu_N = M_\chi M_N / (M_\chi + M_N)$, where $M_N$ denotes the nuclear mass.  
The WIMP--nucleus cross section at zero momentum transfer is represented by $\sigma_N$, and $F(E_R)$ is the nuclear form factor that accounts for the loss of coherence at finite momentum transfer.  
The detector efficiency is assumed to be unity above the threshold.

The minimum WIMP velocity required to produce a recoil of energy $E_R$ is
\begin{equation*}
v_{\min} = \sqrt{\frac{M_N E_R}{2 \mu_N^2}}.
\end{equation*}
The astrophysical dependence of the rate enters through the mean inverse speed,
\begin{equation*}
\eta(v_{\min}) = 
\int_{v_{\min}}^{v_{\rm esc}} 
\frac{f(\mathbf{v} + \mathbf{v}_E)}{v} \, d^3v,
\end{equation*}
where $f(\mathbf{v})$ is the WIMP velocity distribution in the Galactic rest frame, 
$\mathbf{v}_E$ is the Earth’s velocity relative to the Galactic frame, and 
$v_{\rm esc}$ is the Galactic escape velocity.  
This integral represents the fraction of WIMPs with sufficient speed to induce a recoil of energy $E_R$. In the standard halo model~\cite{Standard_halo_model}, the Galactic WIMP velocity distribution is assumed to follow a truncated Maxwell--Boltzmann form with parameters, 
$v_{\rm esc} = 544~\mathrm{km/s}$ (escape velocity), and 
$v_E = 232~\mathrm{km/s}$ (Earth’s velocity). 
Analytic approximations follow the formalism of Lewin and Smith~\cite{WIMP_rate_Lewin}.

\subsection{Spin-Independent WIMP--Nucleon Scattering in Germanium}

We focus on studying the neutrino floor for germanium-based detectors.  
Naturally occurring germanium isotopes are predominantly even--even nuclei with negligible total spin, suppressing spin-dependent (SD) scattering. Consequently, the recoil spectrum is dominated by coherent spin-independent (SI) interactions, allowing SD contributions to be neglected in sensitivity projections.

The WIMP--nucleus cross section at zero momentum transfer is related to the fundamental SI WIMP--nucleon cross section $\sigma_{\chi-n}^{\mathrm{SI}}$ by
\begin{equation}\label{sigma_Nucleus_sigma_nucleon}
\sigma_N = \sigma_{\chi-n}^{\mathrm{SI}} \, A^2 
\left( \frac{\mu_N}{\mu_n} \right)^2,
\end{equation}
where $\mu_n = M_\chi m_n / (M_\chi + m_n)$ is the WIMP--nucleon reduced mass and $m_n$ is the nucleon mass.  
Coherent scattering off all $A$ nucleons enhances the cross section by a factor of $A^2$, while the reduced-mass ratio $(\mu_N / \mu_n)^2$ accounts for the kinematic scaling between the nucleus and the nucleon.

Substituting Eq.~\eqref{sigma_Nucleus_sigma_nucleon} into Eq.~\eqref{eq:dRdE_sigmaN_journal} gives the differential recoil rate per unit detector mass:
\begin{equation}
\frac{dR}{dE_R} = 
\frac{N_A \, \rho_0 \, \sigma_{\chi-n}^{\mathrm{SI}}}{2 A M_\chi \, \mu_n^2} 
A^2 \, F^2(E_R) \, \eta(v_{\min}).
\label{eq:dRdE_sigma_n}
\end{equation}

The nuclear form factor is modeled using the standard Helm parameterization~\cite{WIMP_rate_Lewin}:
\begin{equation}
F^2(E_R) = 
\left[ \frac{3 j_1(q r_n)}{q r_n} \right]^2 
\exp(-q^2 s^2),
\end{equation}
where $j_1(x) = (\sin x / x^2 - \cos x / x)$ is the spherical Bessel function of order one,  
$q = \sqrt{2 M_N E_R}$ is the momentum transfer,  
$s = 0.9~\mathrm{fm}$ is the nuclear skin thickness, and  
$r_n = \sqrt{c^2 + \frac{7}{3}\pi^2 a^2 - 5 s^2}$,  
with $c = 1.23 A^{1/3} - 0.60~\mathrm{fm}$ and $a = 0.52~\mathrm{fm}$.

\subsection{Quenching and the Measured WIMP Energy Spectrum}

In germanium-based detectors operated in high-voltage (HV) mode, such as those used by SuperCDMS~\cite{CDMSliteR3_WIMP_PLR_2019}, the total measured phonon signal is dominated by Neganov--Trofimov--Luke (NTL) phonons~\cite{LUKE1,LUKE2}. The total energy of these NTL phonons is proportional to the amount of ionization produced by particle interactions. For nuclear recoils, only a fraction of the deposited energy is converted into ionization, while the remainder goes into the creation of primary phonons. Consequently, the nuclear recoil energy $E_R$ from WIMP scattering does not entirely lead to NTL phonon production.

This reduced ionization response, known as the \emph{quenching effect}, reflects the lower ionization yield of nuclear recoils relative to electron recoils of the same energy.  
The ionization yield is commonly described by the Lindhard model~\cite{lindhard_quenching_1963}:
\begin{equation}\label{lindhard quenching}
Y(E_R) = \frac{k\, g(\epsilon)}{1 + k\, g(\epsilon)},
\end{equation}
where 
$g(\epsilon) = 3\epsilon^{0.15} + 0.7\epsilon^{0.6} + \epsilon$ and 
$\epsilon = 11.5\,E_R\,Z^{-7/3}$ is the dimensionless energy parameter.  
The constant $k$ is material-dependent.  
For germanium ($Z = 32$), recent CE$\nu$NS results from the CONUS+ Collaboration~\cite{CONUS_reactor_anti-neutrino_observation_Ackermann:2025obx} indicate that $k \approx 0.162$ provides good agreement down to recoil energies of $\sim150~\mathrm{eV_{ee}}$.  
We adopt this value and extrapolate it below $150~\mathrm{eV_{ee}}$, assuming that the Lindhard model remains valid in this regime.
Because both WIMP- and CE$\nu$NS-induced interactions produce nuclear recoils, they are expected to exhibit similar quenching behavior. Consequently, the exact choice of quenching model has little impact on the determination of the neutrino floor. However, applying quenching is important for obtaining a realistic prediction of the experimentally observable recoil spectrum.

\begin{figure}[t]
    \centering
    \includegraphics[width=1\linewidth]{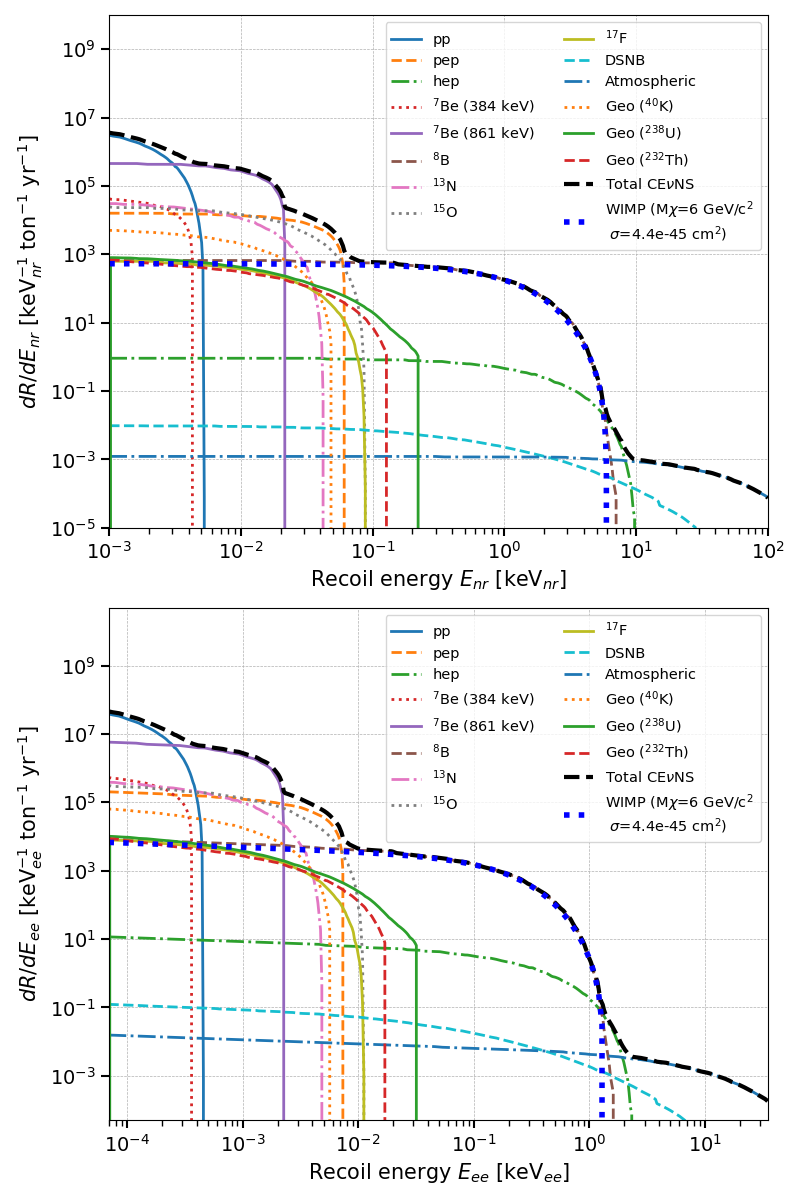}
    \caption{
    Comparison of nuclear-recoil (\textbf{top}) and electron-equivalent (\textbf{bottom}) spectra obtained using the Lindhard quenching model for various CE$\nu$NS components, shown alongside a representative WIMP signal (6~GeV/$c^2$, $\sigma_{\chi-n} = 4.4\times10^{-45}$~cm$^2$).
The $^8$B solar neutrino spectrum closely resembles the WIMP-induced recoil spectrum over the entire energy range of interest, highlighting the challenge of distinguishing low-mass WIMP signals from solar neutrino backgrounds.
}
    \label{fig:CEvNS_WIMP_recoil_spectrum}
\end{figure}

The measured (electron-equivalent) energy \(E_{\mathrm{ee}}\) is related to the recoil energy as
\begin{equation}
E_{\mathrm{ee}} = Y(E_R)\, E_R,
\label{eq:quenching}
\end{equation}
accounting for the quenching effect.  
The differential event rate in measured energy is obtained as
\begin{equation}
\frac{dR}{dE_{\mathrm{ee}}} 
= \frac{dR}{dE_R} 
\frac{dE_R}{dE_{\mathrm{ee}}} 
= \frac{dR}{dE_R} 
\left[\frac{1}{dE_{\mathrm{ee}}/dE_R}\right],
\label{eq:rate_conversion}
\end{equation}
where the derivative \(dE_{\mathrm{ee}}/dE_R\), evaluated using Eq.~\eqref{eq:quenching}, results in the conversion between the recoil and measured energy scales.

The resulting spectrum, \(\frac{dR}{dE_{\mathrm{ee}}}\), represents the observable WIMP-induced signal and forms the basis for sensitivity projections and comparisons with CE$\nu$NS-induced backgrounds.

Figure~\ref{fig:CEvNS_WIMP_recoil_spectrum} shows the differential event rate for a WIMP with mass 6~GeV/$c^2$ and spin-independent cross section \(4.4 \times 10^{-45}~\mathrm{cm}^2\), alongside the CE$\nu$NS background contributions from the various sources discussed in Sec.~\ref{sec:CEvNS_background_modeling}.

\section{CE$\nu$NS-Induced Background Modeling \label{sec:CEvNS_background_modeling}}

\subsection{Differential Scattering Cross Section}

Coherent elastic neutrino--nucleus scattering (CE$\nu$NS) occurs when a neutrino scatters elastically off an entire nucleus at low momentum transfer, resulting in a coherent enhancement of the cross section proportional to the square of the weak nuclear charge~\cite{freedman}. The process is neutrino–flavor independent, as it is mediated by the neutral-current interaction.
Within the Standard Model, the differential cross section for a neutrino of energy $E_\nu$ scattering off a nucleus of mass $M$ and producing a recoil energy $E_R$ can be expressed as~\cite{CEvNS_Crosssection_Scholberg:2005qs}:
\begin{equation}
\frac{d\sigma(E_\nu, E_R)}{dE_R}
 = \frac{G_F^2\,M}{4\pi}\,Q_W^2
 \left(1 - \frac{M E_R}{2 E_\nu^2}\right)
 F^2(E_R),
\end{equation}
where $G_F$ is the Fermi coupling constant, $Q_W = N - (1 - 4\sin^2\theta_W)Z$ is the weak nuclear charge, $N$ and $Z$ are the neutron and proton numbers of the target nucleus, and $\theta_W$ is the weak mixing angle.  
The prefactor associated with $Z$ is close to zero; therefore, the CE$\nu$NS cross section is dominated by the neutron term, leading to a coherent enhancement approximately proportional to $N^2$.  
The kinematic maximum recoil energy is given by
\begin{equation}
E_R^{\rm max} = \frac{2 E_\nu^2}{M + 2 E_\nu}.
\end{equation}

\subsection{Quenching and the Measured CE$\nu$NS Energy Spectrum}

The differential CE$\nu$NS event rate per unit detector mass is expressed as a sum over the differential rates from all neutrino sources:
\begin{equation}
\frac{dR}{dE_R} = \frac{N_A}{A} \sum_j \int_{E_\nu^{\rm min}(E_R)}^{E_\nu^{\rm max,j}} 
\phi_j(E_\nu)\, \frac{d\sigma(E_\nu, E_R)}{dE_R}\, dE_\nu,
\label{eq:diffRatePerMassSum}
\end{equation}
where $\phi_j(E_\nu)$ is the neutrino flux from source $j$,  
$E_\nu^{\rm min}(E_R) = \sqrt{M E_R / 2}$ is the minimum neutrino energy required to produce a nuclear recoil $E_R$, and $E_\nu^{\rm max,j}$ is the endpoint energy of the corresponding spectrum.  
The detector efficiency is assumed to be unity above threshold.  
The various neutrino sources contributing to the CE$\nu$NS background relevant to the neutrino floor are discussed in Sec.~\ref{subsec:CEvNS_sources}.

In cryogenic germanium detectors, only a fraction of the nuclear recoil energy $E_R$ is converted into measurable phonon signals through the Neganov--Trofimov--Luke effect.  
This energy suppression, referred to as the quenching effect, has been described in Sec.~\ref{sec:WIMP_signal_modeling}.  
The experimentally observable spectrum is obtained by folding the differential recoil rate with the quenching factor using Eq.~\ref{eq:rate_conversion}, yielding the measured CE$\nu$NS energy distribution.

\subsection{Sources of CE$\nu$NS Backgrounds}\label{subsec:CEvNS_sources}

CE$\nu$NS-induced backgrounds originate from both natural and man made neutrino sources.  

\subsubsection{Non-Reactor Neutrino Sources}

\paragraph*{Solar Neutrinos.}
Solar neutrinos are primarily produced through the $pp$ chain and the CNO cycle in the Sun.  
They dominate the neutrino flux below $\sim$18~MeV and constitute the primary CE$\nu$NS background for low-mass WIMP searches ($\lesssim 10$~GeV/$c^2$).  
In this study, we include contributions from $pp$, $pep$, $hep$, $^7$Be, $^8$B, $^{13}$N, $^{15}$O, and $^{17}$F neutrinos, using the differential event rates in nuclear recoil energy for a germanium detector from Ref.~\cite{Billard:2013qya}.

\paragraph*{Atmospheric Neutrinos.}
Atmospheric neutrinos are produced by interactions of cosmic rays with Earth's atmosphere and include both neutrinos and antineutrinos.  
These neutrinos are highly energetic, with energies extending up to the TeV scale.  
However, as the neutrino energy increases, the coherence of the CE$\nu$NS process diminishes due to the increasing momentum transfer.  
Following Ref.~\cite{Billard:2013qya}, we consider atmospheric neutrinos up to 1~GeV, which primarily obscure WIMP signals in the mass range 100~GeV/$c^2$–$\mathcal{O}$(10~TeV/$c^2$), and adopt the corresponding differential event rate in nuclear recoil energy for germanium.

\paragraph*{Geoneutrinos.}
Geoneutrinos originate from the $\beta$-decays of long-lived isotopes such as uranium, thorium, and potassium within the Earth’s crust and mantle.  
Their flux exhibits moderate spatial variation due to differences in crustal composition and thickness.  
The geoneutrino flux is typically two orders of magnitude lower than the solar neutrino flux; therefore, its contribution to the CE$\nu$NS background is subdominant.  
Geoneutrinos produce low-energy recoils, primarily affecting WIMP sensitivity near $m_\chi \sim 1$~GeV/$c^2$.  
Given the moderate site-to-site variations and the overall lower flux compared to solar neutrinos, we adopt the differential CE$\nu$NS event rate in germanium detectors in terms of nuclear recoil energy evaluated for the China Jinping Underground Laboratory (CJPL) as a representative case~\cite{Geo_neutrino_Gelmini:2018gqa}. Appendix~\ref{app:geoneutrino_variation} shows that reducing the geoneutrino flux by a factor of two, roughly representing SNOLAB levels~\cite{geo_neutrino_SNOLAB_vs_CJPL,Geo_neutrino_Gelmini:2018gqa}, modifies the no-reactor neutrino floor contributions by at most $\sim$10\% near $M_\chi \simeq 1.7$~GeV/$c^2$. Therefore, the reactor antineutrino flux, which scales with the reactor–detector distance, quickly dominates site-dependent variations, and a single representative geoneutrino model suffices for this study.

\paragraph*{Diffuse Supernova Neutrinos.}
The diffuse supernova neutrino background (DSNB) arises from the cumulative flux of neutrinos emitted by all past core-collapse supernovae in the universe.  
These neutrinos have typical energies of a few MeV, with the spectrum peaking around 3--8~MeV, making them relevant to WIMP searches at intermediate masses (\(\sim 20~\mathrm{GeV}/c^2\)).  
We adopt the DSNB flux and the corresponding differential CE$\nu$NS event rate in germanium detectors as a function of nuclear recoil energy from Ref.~\cite{Billard:2013qya}.

All CE$\nu$NS recoil spectra are converted to the electron-equivalent energy scale using the same Lindhard quenching model applied to WIMP recoils.  
This allows a direct comparison between WIMP-induced and neutrino-induced signals in germanium detectors.  
The differential event rates from individual CE$\nu$NS sources in a germanium detector are shown in Fig.~\ref{fig:CEvNS_WIMP_recoil_spectrum}.  
As evident from the figure, the differential rate from the $^8$B solar neutrino component closely resembles the spectrum of a WIMP with mass 6~GeV/$c^2$ and spin-independent cross section \(4.4 \times 10^{-45}~\mathrm{cm}^2\), highlighting the difficulty of distinguishing low-mass WIMP signals from neutrino-induced backgrounds.

Solar neutrinos dominate the CE$\nu$NS background at low WIMP masses (\(<10~\mathrm{GeV}/c^2\)), while atmospheric neutrinos dominate at higher masses (\(>100~\mathrm{GeV}/c^2\)).  
The fractional uncertainties associated with each neutrino source are summarized in Table~\ref{tab:flux_uncertainties}.  
The uncertainty in the CONUS+ reported quenching parameter ($k$=0.162 $\pm$ 0.004) is on the order of 1\%~\cite{COUNSplus_CEvNS}, which is much smaller than the overall systematic uncertainties in the CE$\nu$NS rate. Since both CE$\nu$NS and WIMP-induced spectra are subject to similar quenching effects, the impact of quenching uncertainty on the neutrino floor determination is expected to be minimal. We evaluated this systematic impact by varying \(k\) within its \(3\sigma\) uncertainty range and found that the resulting neutrino floor remain effectively unchanged. Consequently, we neglect the propagation of quenching-parameter uncertainty in our main calculations. Further discussion is provided in Appendix~\ref{app:quenching_effect}.

\begin{table}[t]
\caption{Fractional uncertainties $\sigma_j$ adopted for each CE$\nu$NS flux component.}
\begin{ruledtabular}
\begin{tabular}{lc}
Component & Uncertainty $\sigma_j$ \\
\hline
pp & 0.6\%~\cite{Overcome_neutrino_floor:2020lva} \\
pep & 1\%~\cite{Overcome_neutrino_floor:2020lva} \\
hep & 30\%~\cite{Overcome_neutrino_floor:2020lva} \\
$^7$Be I (384~keV) & 6\%~\cite{Overcome_neutrino_floor:2020lva} \\
$^7$Be II (861~keV) & 6\%~\cite{Overcome_neutrino_floor:2020lva} \\
$^8$B & 2\%~\cite{Overcome_neutrino_floor:2020lva} \\
$^{13}$N & 15\%~\cite{Overcome_neutrino_floor:2020lva} \\
$^{15}$O & 17\%~\cite{Overcome_neutrino_floor:2020lva} \\
$^{17}$F & 20\%~\cite{Overcome_neutrino_floor:2020lva} \\
DSNB & 50\%~\cite{Overcome_neutrino_floor:2020lva} \\
Atmospheric & 25\%~\cite{Overcome_neutrino_floor:2020lva} \\
Geo ($^{40}$K) & 6\%~\cite{Geo_neutrino_Gelmini:2018gqa} \\
Geo ($^{238}$U) & 30\%~\cite{Geo_neutrino_Gelmini:2018gqa} \\
Geo ($^{232}$Th) & 30\%~\cite{Geo_neutrino_Gelmini:2018gqa} \\
Reactor & 8\%~\cite{Overcome_neutrino_floor:2020lva} \\
\end{tabular}
\end{ruledtabular}
\label{tab:flux_uncertainties}
\end{table}

\subsubsection{Reactor Antineutrinos} \label{subsec:reactor_antineutrino}

Nuclear reactors are intense sources of $\bar{\nu}_e$ emitted in the $\beta$-decays of fission fragments from $^{235}$U, $^{238}$U, $^{239}$Pu, and $^{241}$Pu.  
These low-energy antineutrinos produce sub-keV recoils via CE$\nu$NS, overlapping the expected signal region for light WIMPs in germanium detectors.  

To evaluate the reactor-induced CE$\nu$NS background, we consider a representative reactor similar to the Leibstadt (KKL) plant in Switzerland~\cite{conus+_experiment_2024}, operating at a thermal power of 3.6~GW$_{\text{th}}$.  
The reactor antineutrino flux at a detector located a distance $L$ from the reactor core is
\begin{equation}
\phi_{\bar{\nu}_e}(E_\nu) = \frac{P_{\rm th}}{4 \pi L^2 \langle E_f \rangle} \sum_i f_i \, S_i(E_\nu),
\end{equation}
where $P_{\rm th}$ is the reactor thermal power, $\langle E_f \rangle \approx 202.5$~MeV is the average energy released per fission~\cite{energy_per_fission}, $f_i$ is the fission fraction of isotope $i$, and $S_i(E_\nu)$ is the emitted $\bar{\nu}_e$ spectrum per fission.  
The flux scales as $1/L^2$, making distance a critical parameter for near-core experiments.  
The fission fractions are listed in Table~\ref{tab:fission_fractions}.

\begin{table}[h!]
\centering
\caption{Used fission fractions for a  gigawatt-scale reactor, based on typical values for the KKL reactor~\cite{conus+_experiment_2024}.}
\label{tab:fission_fractions}
\begin{tabular}{lc}
\hline
Isotope & Fission Fraction $f_i$ \\
\hline
$^{235}$U & 53\% \\
$^{239}$Pu & 32\% \\
$^{238}$U & 8\% \\
$^{241}$Pu & 7\% \\
\hline
\end{tabular}
\end{table}

The total reactor spectrum combines a parameterized model above 2~MeV with tabulated data below 2~MeV.  
For $E_\nu > 2$~MeV, the spectra are parameterized as~\cite{Hubber_reactor_spectrum_PhysRevC.84.024617,CONNIE_anti_neutrino_spectrum:2019xid}
\begin{equation}
\frac{dN_{\bar{\nu}_e}}{dE_{\bar{\nu}_e}} = a \exp(a_0 + a_1 E_{\bar{\nu}_e} + a_2 E_{\bar{\nu}_e}^2),
\end{equation}
with coefficients listed in Table~\ref{tab:param_highE}.

\begin{table}[h!]
\centering
\caption{Parameterization constants for $\bar{\nu}_e$ spectra  per MeV per fission above 2~MeV.}
\label{tab:param_highE}
\begin{tabular}{ccccc}
\hline
Isotope & $a$ & $a_0$ & $a_1$ & $a_2$ \\
\hline
$^{235}$U & 1.0461 & 0.870 & –0.160 & –0.0910 \\
$^{238}$U & 1.0719 & 0.976 & –0.162 & –0.0790 \\
$^{239}$Pu & 1.0527 & 0.896 & –0.239 & –0.0981 \\
$^{241}$Pu & 1.0818 & 0.793 & –0.080 & –0.1085 \\
\hline
\end{tabular}
\end{table}

Below 2~MeV, the spectra are taken from tabulated data (Table~\ref{tab:lowE_flux}):

\begin{table}[h!]
\centering
\caption{Antineutrino spectra below 2~MeV (in $\bar{\nu}_e$/MeV/fission)~\cite{CONNIE_anti_neutrino_spectrum:2019xid}.}
\label{tab:lowE_flux}
\begin{tabular}{c|cccc}
\hline
$E_\nu$ [MeV] & $^{235}$U & $^{239}$Pu & $^{238}$U & $^{241}$Pu \\
\hline
0.008 & 0.024 & 0.140 & 0.089 & 0.200 \\
0.016 & 0.092 & 0.560 & 0.350 & 0.790 \\
0.031 & 0.350 & 2.130 & 1.320 & 3.000 \\
0.062 & 0.610 & 0.640 & 0.650 & 0.590 \\
0.125 & 1.980 & 1.990 & 2.020 & 1.850 \\
0.250 & 2.160 & 2.080 & 2.180 & 2.140 \\
0.500 & 2.660 & 2.630 & 2.910 & 2.820 \\
0.750 & 2.660 & 2.580 & 2.960 & 2.900 \\
1.000 & 2.410 & 2.320 & 2.750 & 2.630 \\
1.500 & 1.690 & 1.480 & 1.970 & 1.750 \\
2.000 & 1.260 & 1.080 & 1.500 & 1.320 \\
\hline
\end{tabular}
\end{table}

\begin{figure}
    \centering
    \includegraphics[width=1\linewidth]{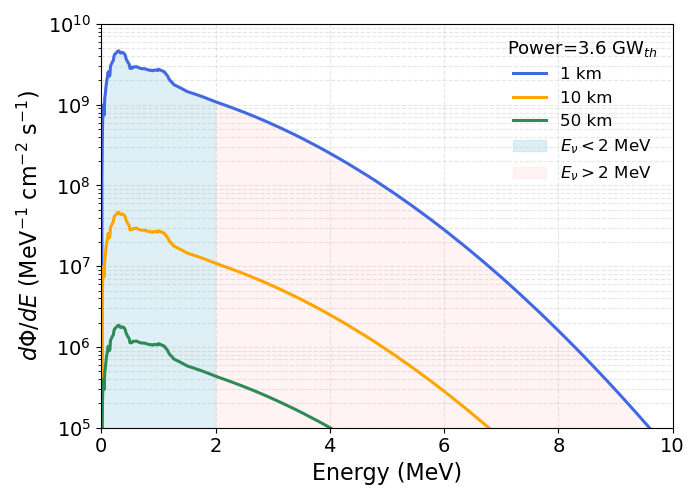}
    \caption{Reactor antineutrino flux at 1, 10, and 50~km from a 3.6~GW$_\mathrm{th}$ reactor core.
The total spectrum combines a parameterized model above 2~MeV with tabulated data below, as described in Sec.~\ref{subsec:reactor_antineutrino}.
The overall flux decreases with distance following the expected $1/L^2$ dependence.}
    \label{fig:reactor_anti_neutrino_flux}
\end{figure}

At low energies (around 1~MeV and below), an additional contribution to the $\bar{\nu}_e$ flux arises from neutron capture on $^{238}$U, with the corresponding spectrum taken from Ref.~\cite{TEXONO_U238_capture:2006xds}.  
This spectrum was normalized and scaled according to its antineutrino yield per fission, taken as 1.2~\cite{CONNIE_anti_neutrino_spectrum:2019xid}.  
The combined reactor antineutrino spectra at different reactor--detector distances are shown in Fig.~\ref{fig:reactor_anti_neutrino_flux}. Following Ref.~\cite{Overcome_neutrino_floor:2020lva}, we adopt a nominal reactor antineutrino flux uncertainty of 8\%. This value is broadly in line with the $\sim$7.8\% deficit reported by Daya Bay in the observed versus predicted yield of the $^{235}$U fission isotope~\cite{DayaBay_reactor_flux_uncertainty}.

\begin{figure*}[t]
    \centering
    \includegraphics[width=0.45\linewidth]{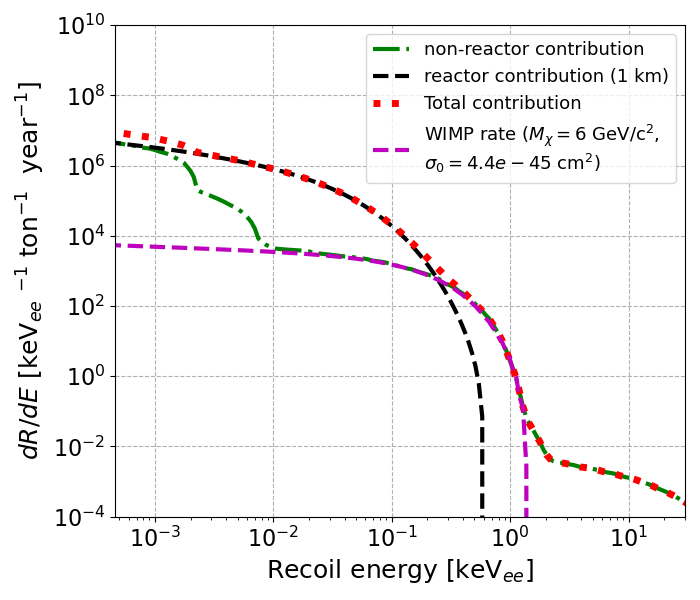}
    \includegraphics[width=0.45\linewidth]{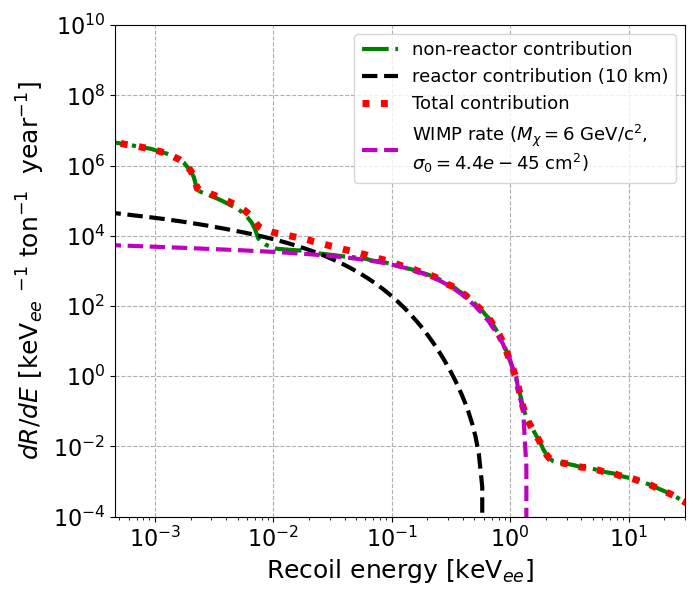}
    \includegraphics[width=0.45\linewidth]{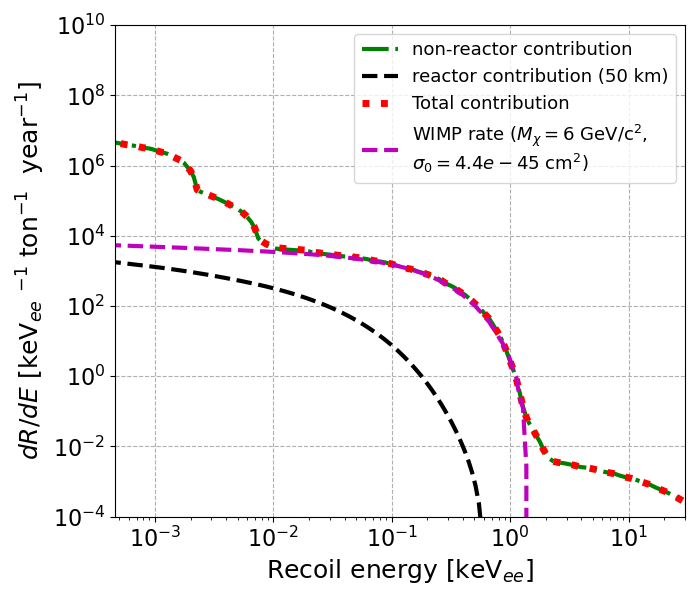}
    \caption{Comparison of differential CE$\nu$NS event rates from reactor antineutrinos and non-reactor (solar, atmospheric, DSNB, and geoneutrino) sources.
Results are shown for detector locations at 1~km (\textbf{upper left}), 10~km (\textbf{upper right}), and 50~km (\textbf{bottom}) from a 3.6~GW$_\mathrm{th}$ reactor core.
The reactor contribution begins to dominate over non-reactor backgrounds at reactor–detector separations below approximately 10~km.}
    \label{fig:reactor_diff_rate}
\end{figure*}

Reactor antineutrinos can constitute an important component of the CE$\nu$NS background for direct detection WIMP search experiments, especially for sites located near nuclear power plants.  
Their contribution strongly depends on the distance between the detector and the reactor core, leading to large variations in the expected event rates.  
As shown in Fig.~\ref{fig:reactor_diff_rate}, the CE$\nu$NS event rate can vary by a few orders of magnitude for detector locations within $\sim$50~km of a reactor and begins to surpass the solar neutrino background at baselines below $\sim$10~km.  
Accurate modeling of the site-dependent reactor flux is therefore essential for evaluating the neutrino floor and assessing the discovery reach of future low-threshold dark matter experiments.

We validated our reactor antineutrino CE$\nu$NS recoil spectrum against the CE$\nu$NS spectrum measured by the CONUS+ experiment~\cite{CONUS+_first_observation_2025}.
As discussed in Appendix~\ref{app:reactor_CEnuNS_validation}, our Standard Model prediction for the CE$\nu$NS recoil spectrum is in good agreement with the experimental data from CONUS+, measured at the same KKL reactor.

Since this study focuses on the dependence of reactor antineutrino backgrounds on distance (from a few to about 100~km), we assume that non-reactor neutrino flux components and their associated event rates remain constant when varying the detector location.

\section{Impact of Reactor Antineutrinos on the Neutrino Floor\label{sec:neutrino_floor_definition}}

The \emph{neutrino floor} defines the sensitivity threshold of direct WIMP search experiments, below which CE$\nu$NS backgrounds become statistically indistinguishable from potential WIMP signals.  
In this regime, uncertainties in the neutrino background limit the achievable discovery significance, such that further increases in exposure yield diminishing gains in sensitivity.

In this work, we examine two widely used formulations of the neutrino floor and assess the impact of reactor antineutrinos:  
the \emph{discovery-limit} approach proposed by Billard \textit{et al.}~\cite{Billard:2013qya}, and  
the \emph{opacity-based} or \emph{neutrino-fog} framework introduced by O’Hare~\cite{OHare:2021utq}.  

\subsection{Discovery-Limit Based Neutrino Floor}

In the discovery-limit framework~\cite{Billard:2013qya}, the neutrino floor is defined as the minimum WIMP-nucleon cross section for which a WIMP signal can be detected with at least $3\sigma$ significance in 90\% of hypothetical experiments, in the presence of CE$\nu$NS backgrounds.

The detection significance is evaluated using a binned profile-likelihood ratio test, following standard methods for new-physics searches~\cite{cowan2011asymptotic}.  
Event counts in each nuclear-recoil energy bin are assumed to follow independent Poisson distributions, while uncertainties in individual neutrino flux components are treated as Gaussian nuisance parameters.  
The likelihood function is written as
\begin{align}
\mathcal{L}(\boldsymbol{\alpha}) 
&= \prod_{i=1}^{N_\mathrm{bins}} 
    \frac{\!\big(\sum_j \alpha_j B_{ij} + S_i\big)^{n_i}}{n_i!} 
    e^{-\sum_j \alpha_j B_{ij} - S_i}  \nonumber\\[3pt]
&\quad \times 
    \prod_j \exp\!\left[-\frac{(\alpha_j - 1)^2}{2\sigma_j^2}\right],
\end{align}
where $n_i$ is the observed number of events in bin $i$,  
$S_i$ is the predicted WIMP signal for a given dark matter mass and cross section,  
$B_{ij}$ is the CE$\nu$NS background contribution from flux component $j$,  
$\alpha_j$ is the normalization parameter scaling the $j$th flux component,  
and $\sigma_j$ denotes its fractional uncertainty as provided in Tab.~\ref{tab:flux_uncertainties}.

The corresponding conditional (background-only) and unconditional (signal-plus-background) log-likelihoods are
\begin{widetext}
\begin{align}
\ln \mathcal{L}_\mathrm{cond} 
&= \sum_i n_i \ln\!\left(\sum_j \alpha_j b_{ij}\right)
   - \sum_i\sum_j \alpha_j b_{ij} - \sum_j \frac{(\alpha_j-1)^2}{2\sigma_j^2}, \\[4pt]
\ln \mathcal{L}_\mathrm{uncond} 
&= \sum_i n_i \ln\!\left(s_i + \sum_j \alpha_j b_{ij}\right)
   - \sum_i \!\left(s_i + \sum_j \alpha_j b_{ij}\right) - \sum_j \frac{(\alpha_j-1)^2}{2\sigma_j^2}.
\end{align}
\end{widetext}

The profile-likelihood ratio test statistic is defined as
\begin{equation}
q_0 = 2 \!\left( 
\ln \mathcal{L}_\mathrm{uncond}\!\big|_{\alpha_j = \hat{\alpha}_j}
- \ln \mathcal{L}_\mathrm{cond}\!\big|_{\alpha_j = \hat{\!\hat{\alpha}}_j}
\right),
\end{equation}
where $\hat{\alpha}_j$ are the \emph{unconditional maximum-likelihood estimates} obtained by maximizing $\mathcal{L}_\mathrm{uncond}$, and $\hat{\!\hat{\alpha}}_j$ are the \emph{conditional maximum-likelihood estimates} obtained by maximizing $\mathcal{L}_\mathrm{cond}$ under the background-only hypothesis.

The discovery significance, expressed in Gaussian $\sigma$ units, is
\begin{equation}
Z = \sqrt{q_0},
\end{equation}
which quantifies the significance for rejecting the background-only hypothesis in favor of the signal-plus-background hypothesis.

To determine the \emph{discovery-limit cross section}, we perform 1000 Monte Carlo pseudo-experiments for each assumed WIMP mass and cross section.  
Each pseudo-dataset is generated by applying independent Poisson fluctuations to the expected number of events in each energy bin, based on the combined WIMP signal and CE$\nu$NS background model.  
For each realization, the likelihood ratio $q_0$ and the corresponding significance $Z$ are computed, and the resulting $Z$ distribution captures the expected statistical variation among experiments.

The discovery-limit cross section, $\sigma_{\chi-n}^{\mathrm{disc}}$, is defined as the smallest cross section for which at least 90\% of pseudo-experiments yield $Z \ge 3$.  
In statistical terms, this corresponds to a \emph{power of 90\%} to reject the background-only hypothesis at $3\sigma$ significance, providing a robust and experimentally meaningful detection threshold.

Repeating this procedure across a range of WIMP masses yields the discovery threshold as a function of mass, and the curve connecting these thresholds defines the discovery-limit neutrino floor.

To probe a broad WIMP mass range, we perform two sets of pseudo-experiments with different detector thresholds, following the approach of Ref.~\cite{Billard:2013qya}.  
A low-threshold configuration (5.3~eV$_\mathrm{nr}$ or 0.46~eV$_\mathrm{ee}$ using Lindhard quenching) captures the impact of solar neutrinos on low-mass WIMPs, while a high-threshold setup (7.2~keV$_\mathrm{nr}$ or 1.65~keV$_\mathrm{ee}$) suppresses the solar contribution and focuses on DSNB and atmospheric neutrinos relevant for high-mass WIMPs.  
The most sensitive discovery limit from these two regimes defines the neutrino floor across the full WIMP mass range.

The exposures used in the pseudo-experiments are chosen to provide statistically meaningful sampling of the neutrino background.  
For the low-threshold setup, the exposure corresponds to roughly 200 expected $^8$B solar neutrino events, while the high-threshold configuration yields about 500 total neutrino events~\cite{Ruppin_followup:2014bra,Billard:2013qya}.  
These correspond to exposures of approximately 0.4~ton-year and $1.43\times10^4$~ton-year, respectively.  
With these choices, the statistical uncertainty on the total neutrino background remains smaller than the systematic uncertainty in the neutrino flux, allowing the pseudo-experiments to capture both statistical fluctuations and systematic uncertainties.  

It should be noted that these ultra-low threshold and ultra-high exposure values are somewhat idealized and not yet representative of current detector capabilities. They serve only as mathematical constructs that allow the neutrino floor to be evaluated in the asymptotic, systematics-dominated regime. Prototype detectors such as those developed by SuperCDMS~\cite{HVeVR1,SuperCDMS:HVeV_compton_steps,SuperCDMS:HVeV4_DM} and CRESST~\cite{CRESST_2024_WIMP} have achieved eV-scale thresholds but remain limited to gram-scale target masses, making ton-year exposures currently unfeasible.

All calculations incorporate the flux uncertainties summarized in Table~\ref{tab:flux_uncertainties} as Gaussian priors and include contributions from reactor, astrophysical, and geoneutrino CE$\nu$NS sources.  
This treatment ensures that the projected sensitivity accurately reflects the systematic uncertainties in the neutrino fluxes.

\begin{figure}[t]
    \centering
    \includegraphics[width=1\linewidth]{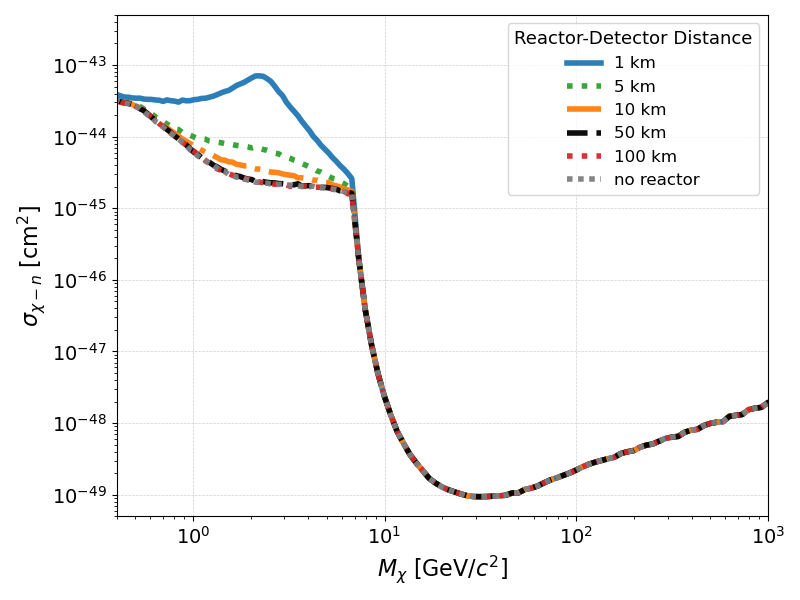}
\caption{Impact of reactor antineutrinos on the discovery-limit neutrino floor for low-mass WIMPs.
Results are shown for detector locations at 1, 5, 10, 50, and 100~km from a 3.6~GW$_\mathrm{th}$ reactor core, along with the no-reactor baseline.}
\label{fig:reactor_distance}
\end{figure}

Figure~\ref{fig:reactor_distance} shows the impact of reactor antineutrinos on the neutrino floor at varying reactor–detector distances.  
The effect is most pronounced for dark matter masses below $\sim$10~GeV/$c^2$, with the largest deviation observed around 2.5--3~GeV/$c^2$.  
These results indicate that proximity to a reactor can substantially elevate the neutrino floor by orders of magnitude for low-mass WIMPs due to the enhanced reactor-induced CE$\nu$NS background.  
As the distance increases, the reactor contribution diminishes and the sensitivity gradually converges toward the baseline neutrino floor.

\begin{figure}
    \centering
    \includegraphics[width=1\linewidth]{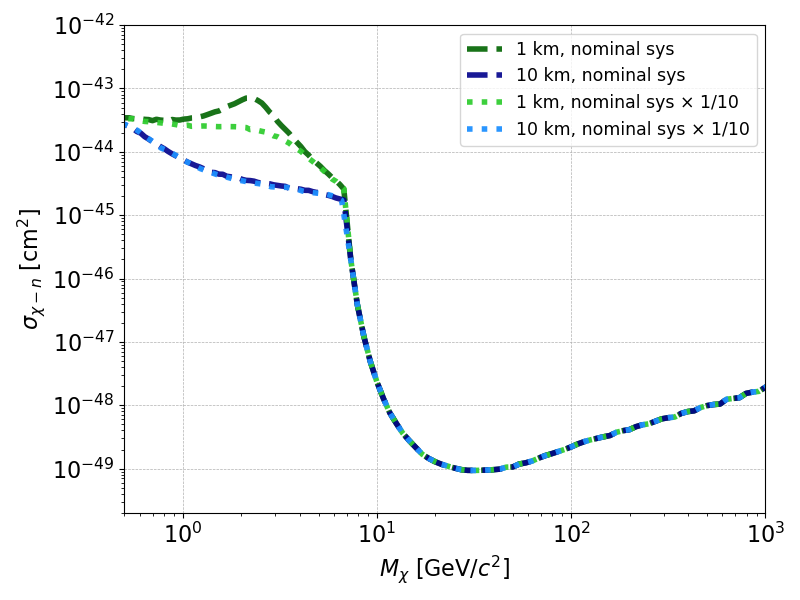}
    \caption{Effect of the discovery-limit–based neutrino floor on reactor antineutrino flux uncertainties.
Results are shown for reactor–detector distances of 1~km and 10~km, assuming a nominal 8\% systematic uncertainty in the reactor antineutrino flux, and compared with a scenario where this uncertainty is reduced by a factor of ten.}
    \label{fig:reactor_systematics}
\end{figure}

We studied how systematic uncertainties in the reactor antineutrino flux influence the neutrino floor.  
Figure~\ref{fig:reactor_systematics} illustrates the impact of varying the systematic uncertainty in the reactor flux for two reactor–detector distances: 1~km and 10~km.  
For the 1~km scenario, a tenfold reduction in flux uncertainty noticeably improves the sensitivity, effectively opening up a larger region of WIMP parameter space accessible to discovery.  
In contrast, for a 10~km separation, the effect of flux uncertainty is much less pronounced due to the reduced reactor contribution.

\begin{figure}
    \centering
    \includegraphics[width=1\linewidth]{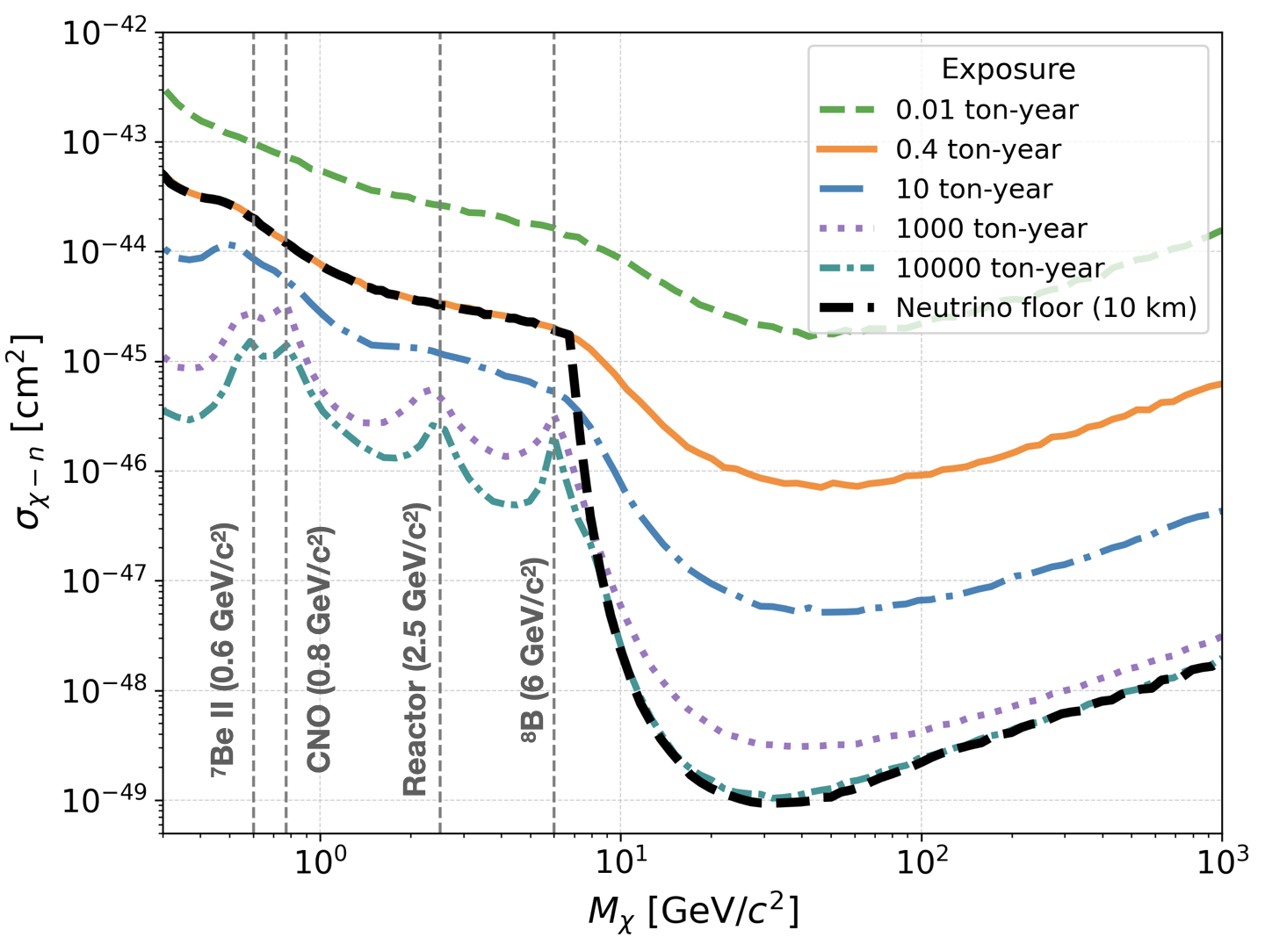}
    \caption{
Dependence of the discovery limit on total exposure for a pseudo-experiment with a nuclear-recoil threshold of 5.3~eV$_\mathrm{nr}$ (equivalent to 0.46~eV$_\mathrm{ee}$ under the Lindhard quenching model). A reactor–detector distance of 10~km is assumed.  
Increasing exposure improves sensitivity to smaller cross sections, although the gain diminishes in the WIMP–mass regime where neutrino-induced recoils become spectrally degenerate with the WIMP signal.  
The threshold and high-exposure combination shown here is idealized and is used to isolate the physics of WIMP–neutrino spectral degeneracy; it is not intended to represent the current or near-future capabilities of germanium detectors.
}
    \label{fig:discovery_threshold_vs_exposure}
\end{figure}

Figure~\ref{fig:discovery_threshold_vs_exposure} illustrates how increasing exposure affects WIMP search sensitivity.  
Higher exposure enhances both the WIMP signal and background counts, reducing statistical uncertainty and improving discrimination between dark matter signals and CE$\nu$NS backgrounds. However, once systematic uncertainties dominate, further exposure yields diminishing returns.  
This limitation is most evident where the WIMP and neutrino recoil spectra are nearly degenerate—for instance, at $M_\chi \sim 6$~GeV/$c^2$, where the WIMP signal resembles the $^8$B solar neutrino spectrum.  
This behavior underscores the intrinsic sensitivity ceiling imposed by neutrino backgrounds.

To overcome the explicit dependence of the discovery-limit-based neutrino floor definition on exposure and threshold, a complementary definition of the neutrino floor based on the concept of \emph{opacity} was introduced. This framework naturally incorporates exposure-dependent effects into the sensitivity definition.

\subsection{Opacity-Based Neutrino Floor}

A more recent definition, introduced by O’Hare~\cite{OHare:2021utq}, characterizes the \emph{neutrino fog} as the transition region where the dark matter signal spectrum becomes progressively obscured by coherent elastic neutrino–nucleus scattering backgrounds, primarily due to systematic uncertainties in neutrino fluxes.
Within this framework, the neutrino floor is defined as the boundary between the Poisson-statistics–dominated regime and the systematic-uncertainty–dominated regime, quantified by the dimensionless \emph{opacity} parameter.
\begin{figure*}
\includegraphics[width=0.49\linewidth]{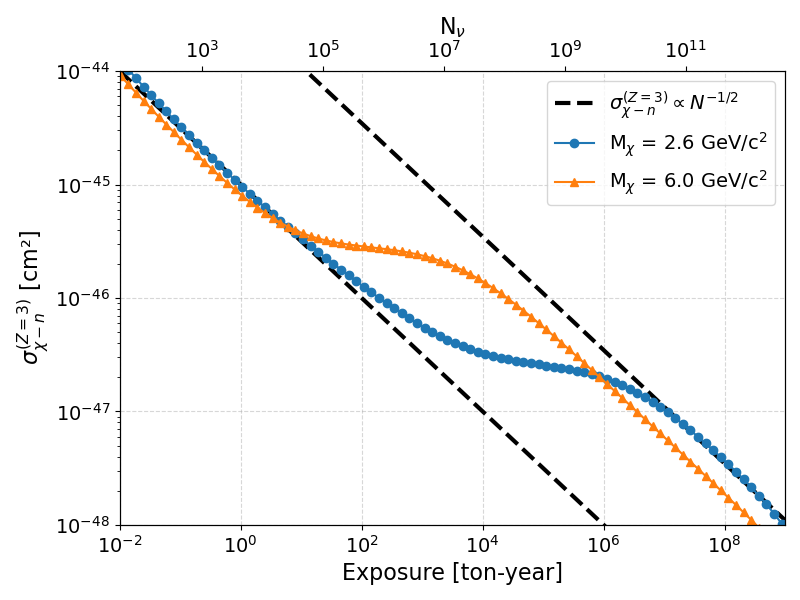}
\includegraphics[width=0.49\linewidth]{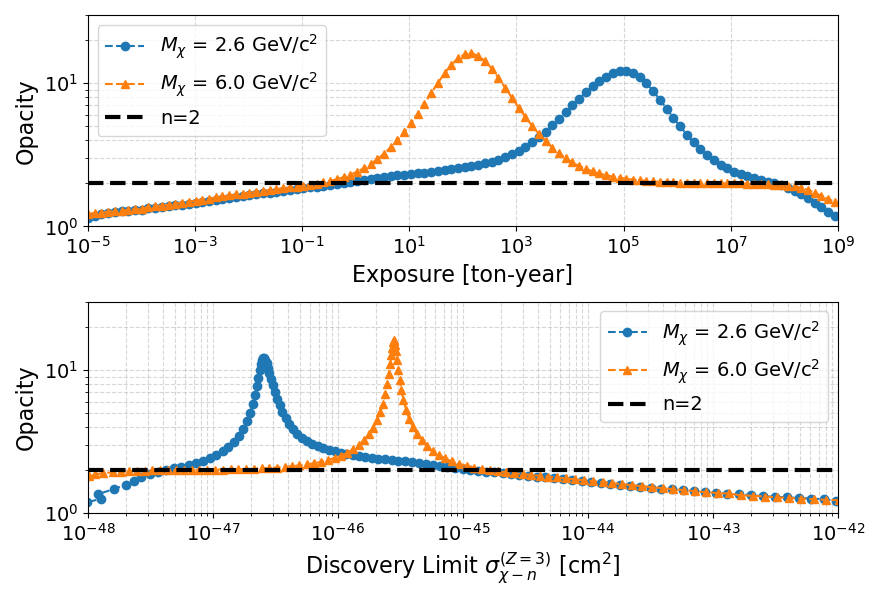}
\caption{
\textbf{Left}: Discovery-limit cross section as a function of exposure for two representative WIMP masses:
(a) 2.6~GeV/$c^2$, where reactor $\bar{\nu}_e$ backgrounds closely mimic the WIMP recoil spectrum (evaluated for a 50~km reactor–detector baseline), and
(b) 6~GeV/$c^2$, where the $^8$B solar neutrino spectrum produces a similar recoil signature.
\textbf{Right}: Dependence of the opacity parameter on detector exposure (top) and discovery-limit cross section (bottom) for the same 50~km baseline, shown for both the WIMP masses.
}
\label{fig:opacity_vs_exposure}
\end{figure*}

At low exposures, discovery sensitivity improves approximately as the inverse square root of exposure, governed by Poisson fluctuations in event counts.  
As exposure increases, further gains become limited (Fig.~\ref{fig:discovery_threshold_vs_exposure}) by systematic uncertainties in the CE$\nu$NS background normalization.

Following methods described in~\cite{OHare:2021utq}, the discovery-limit cross section at fixed detection significance $Z=3$, $\sigma_{\chi-n}^{(Z=3)}$, is evaluated across a grid of WIMP masses $M_\chi$ and exposures.  
For each exposure, the total CE$\nu$NS background $N_\nu$ is obtained by summing contributions from solar, reactor, diffuse supernova, geoneutrino, and atmospheric neutrino sources.  
The \emph{opacity parameter}, $n$, quantifies how the discovery sensitivity degrades with increasing neutrino background (and equivalently with exposure).  It is defined as the inverse logarithmic slope of the discovery-limit cross section with respect to the total CE$\nu$NS background~\cite{OHare:2021utq}:
\begin{equation}
n = -\left( \frac{d \ln \sigma_{\chi-n}^{(Z=3)}}{d \ln N_\nu} \right)^{-1}.
\label{eq:opacity}
\end{equation}

Figure~\ref{fig:opacity_vs_exposure}~(left) shows how the discovery sensitivity, expressed as $\sigma_{\chi\text{--}n}^{(Z=3)}$, evolves with the expected number of neutrino events, revealing distinct scaling behaviors across exposure regimes. At low exposures, $\sigma_{\chi\text{--}n}^{(Z=3)}$ follows the expected $N_\nu^{-1/2}$ dependence, where Poisson fluctuations dominate. As the exposure increases, the improvement in sensitivity gradually saturates, indicating a transition to a regime dominated by systematic uncertainties in the neutrino flux, where further gains are limited and the opacity correspondingly increases. At very high exposures, the scaling returns to the Poisson-dominated behavior, again consistent with the $N_\nu^{-1/2}$ trend. The figure shows these features for two representative WIMP masses: 2.6~GeV/$c^{2}$, where reactor antineutrinos form the dominant CE$\nu$NS background, and 6.0~GeV/$c^{2}$, where $^8$B solar neutrinos majorly contribute. Differences in flux normalization and systematic uncertainties alter the position and extent of the saturation region between the two cases.

The \emph{opacity-based neutrino floor} is defined as the locus in the $(M_\chi, \sigma_{\chi-n})$ plane where the scaling exponent $n=2$,  
marking the transition between the statistics-limited and systematics-limited regimes. The region below this curve marks the regime where increasing exposure yields diminishing returns in sensitivity, as CE$\nu$NS backgrounds increasingly obscure potential WIMP signals—the so-called \emph{neutrino fog}. However, since the WIMP and CE$\nu$NS recoil spectra are not perfectly degenerate, this limitation is not absolute; at sufficiently high exposures, the sensitivity can partially recover the Poisson-like scaling as the spectral differences become statistically distinguishable~\cite{Overcome_neutrino_floor:2020lva,OHare:2021utq}.

\begin{figure*}[t]
\centering
\includegraphics[width=0.45\linewidth]{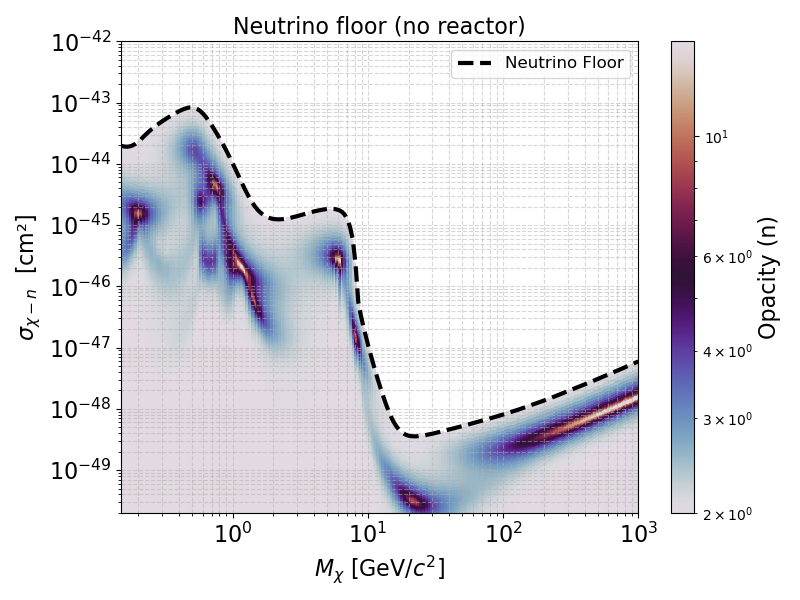}
\includegraphics[width=0.45\linewidth]{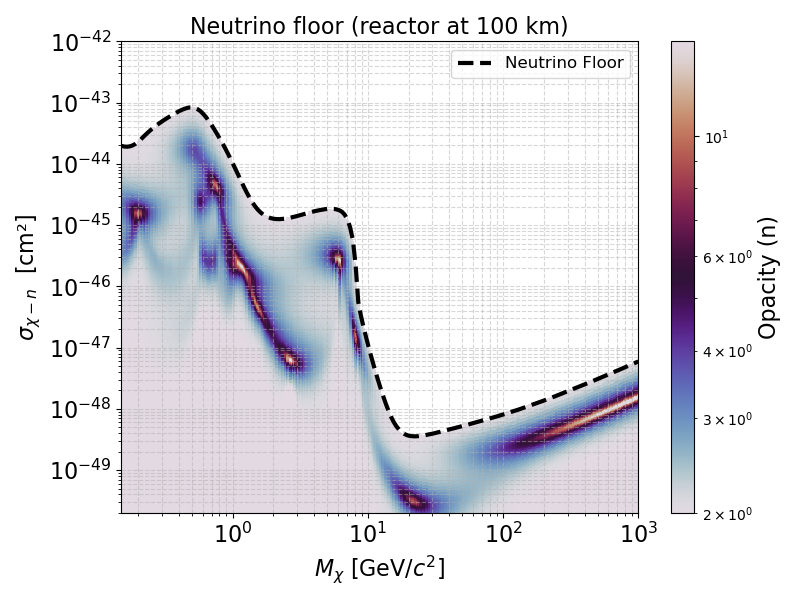}
\includegraphics[width=0.45\linewidth]{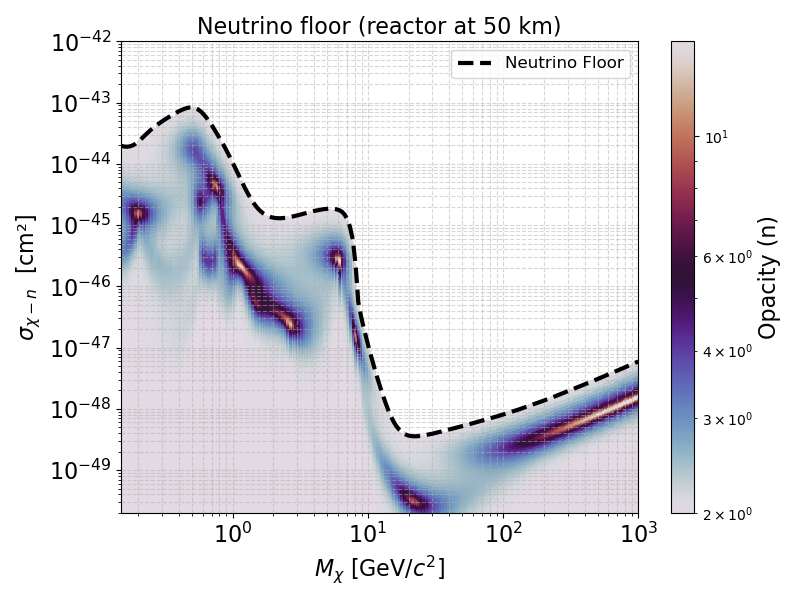}
\includegraphics[width=0.45\linewidth]{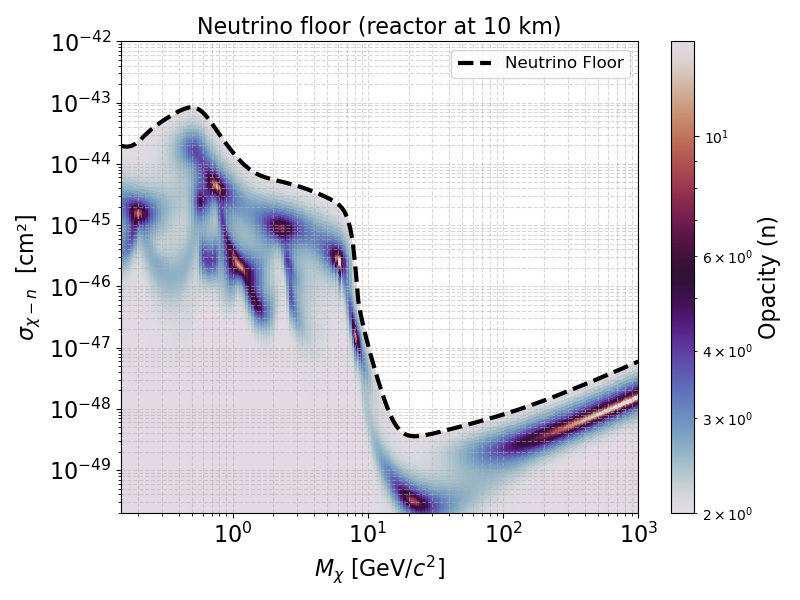}
\includegraphics[width=0.45\linewidth]{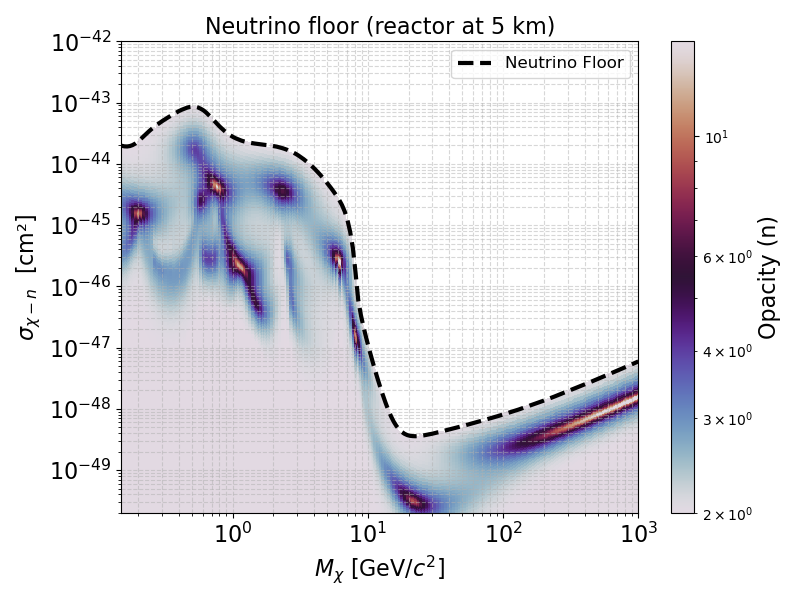}
\includegraphics[width=0.45\linewidth]{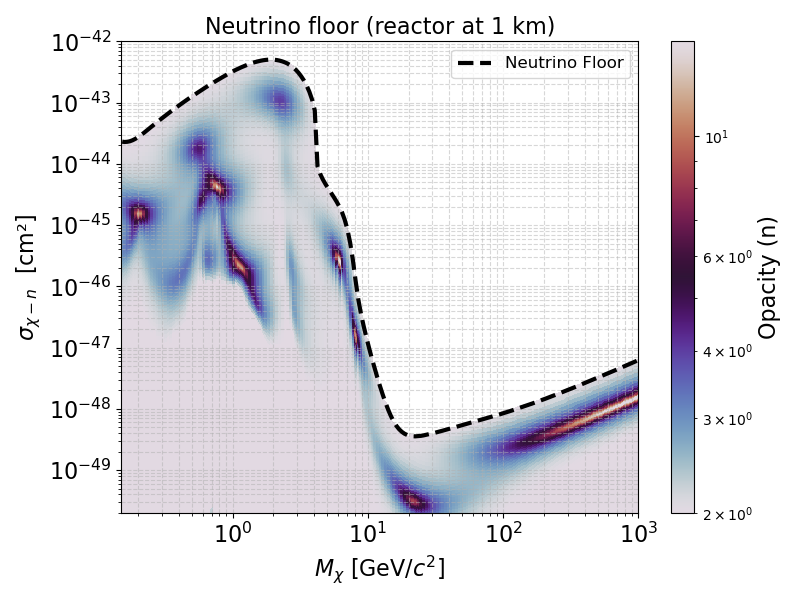}
\caption{
Impact of reactor $\bar{\nu}_e$ flux on the opacity-based neutrino floor for reactor–detector baselines of 1, 5, 10, 50, and 100~km.  
A no-reactor scenario is also shown as the baseline.  
The reactor spectrum predominantly affects WIMP searches around $M_\chi \sim 2.5$–3~GeV/$c^2$.  
As the baseline decreases, the ``foggy'' region of discovery sensitivity shifts upward, with the reactor-induced background overtaking the solar-neutrino contribution at distances below $\sim$10~km. In contrast to the discovery-limit definition, the opacity-based neutrino floor does not depend on threshold and exposure choices in pseudo-experiments.
}
\label{fig:impact_of_reactor_antineutrino_opacity}
\end{figure*}

Figure~\ref{fig:opacity_vs_exposure} (right) shows how the opacity parameter evolves with exposure (top) and discovery limit (bottom) for the two WIMP masses:  
2.6~GeV/$c^2$ and 6.0~GeV/$c^2$.

The resulting neutrino fog at varying reactor–detector distances is shown in Fig.~\ref{fig:impact_of_reactor_antineutrino_opacity},  
alongside the no-reactor baseline.  
In the absence of reactor antineutrinos, the low-mass DM sensitivity is primarily limited by solar neutrinos.  
An additional opaque region appears around $M_\chi \sim 2.5$–3~GeV/$c^2$ due to the contribution from reactor antineutrinos,  
shifting to higher cross sections as the reactor–detector distance decreases.  
For distances shorter than $\sim$10~km, the reactor contribution surpasses the solar component,  
significantly raising the neutrino floor in the low-mass region by up to a few orders of magnitude.  
Unlike the discovery-limit–based floor, the opacity-based definition depends weakly on the absolute exposure choice in pseudo experiments,
making the impact to the baseline floor a more robust physical effect.
These results highlight that reactor antineutrino fluxes constitute a critical, site-dependent background for sub-10~GeV/$c^2$ WIMP searches.  
Careful consideration of reactor proximity is therefore essential when designing low-threshold experiments,  
as even modest reactor contributions can dominate the low-energy CE$\nu$NS spectrum and substantially reduce dark matter discovery sensitivity. 

Appendix~\ref{app:baseline_comparison} presents a comparison of the baseline neutrino floor, i.e., the neutrino floor obtained without including reactor antineutrinos, from this study with previously published results.  
The comparison is shown for both the discovery-limit–based approach~\cite{Billard:APPEC_report} and the opacity-based approach~\cite{OHare:2021utq}, focusing on Ge-based detectors.  
Overall, our baseline limits show reasonable agreement with previously published results under both definitions.

\section{Implications for Future Dark Matter Experiments \label{Impact in next generation experiment}}

\begin{figure*}
    \centering
    \includegraphics[width=0.48\linewidth]{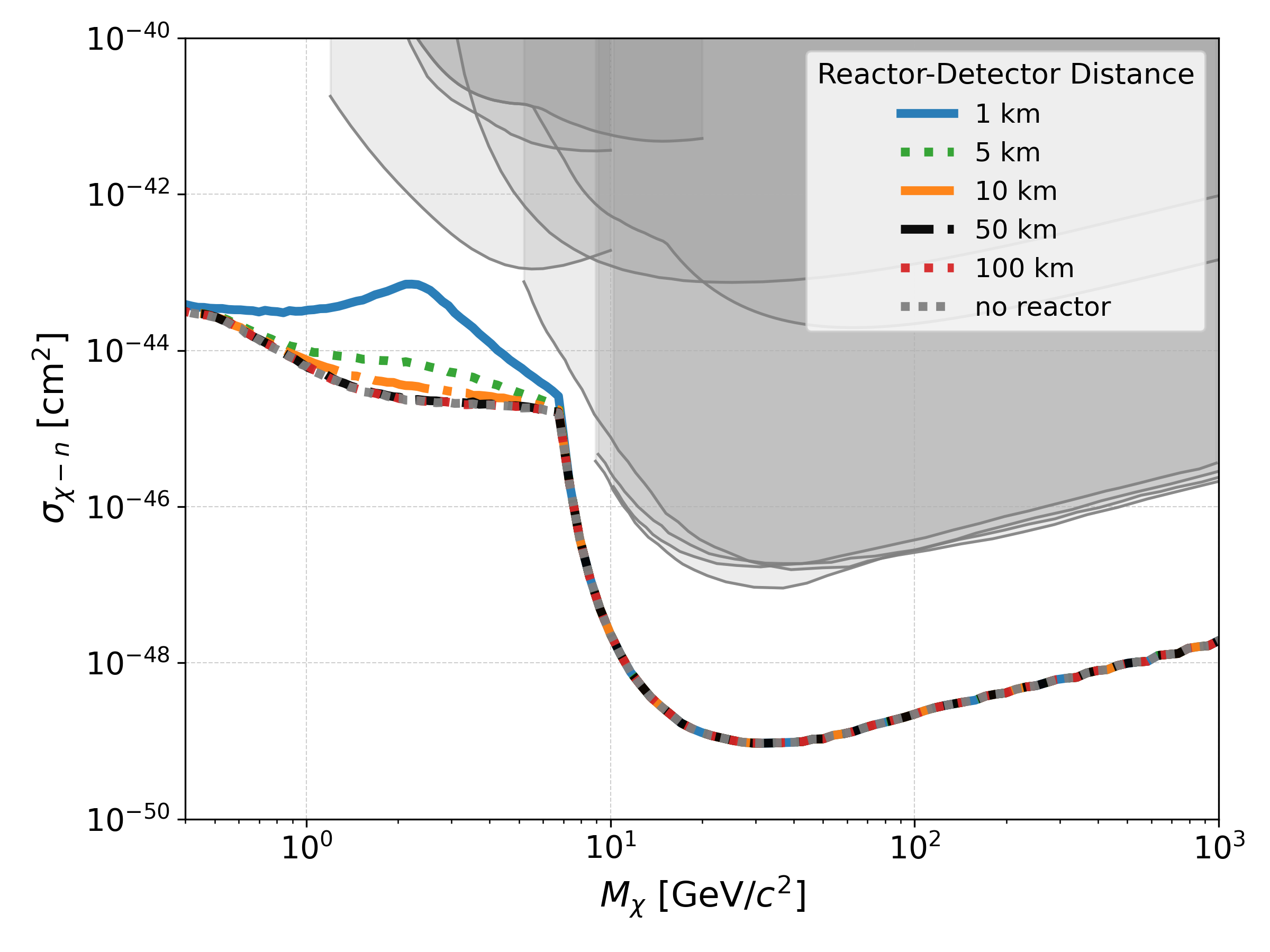}
    \includegraphics[width=0.48\linewidth]{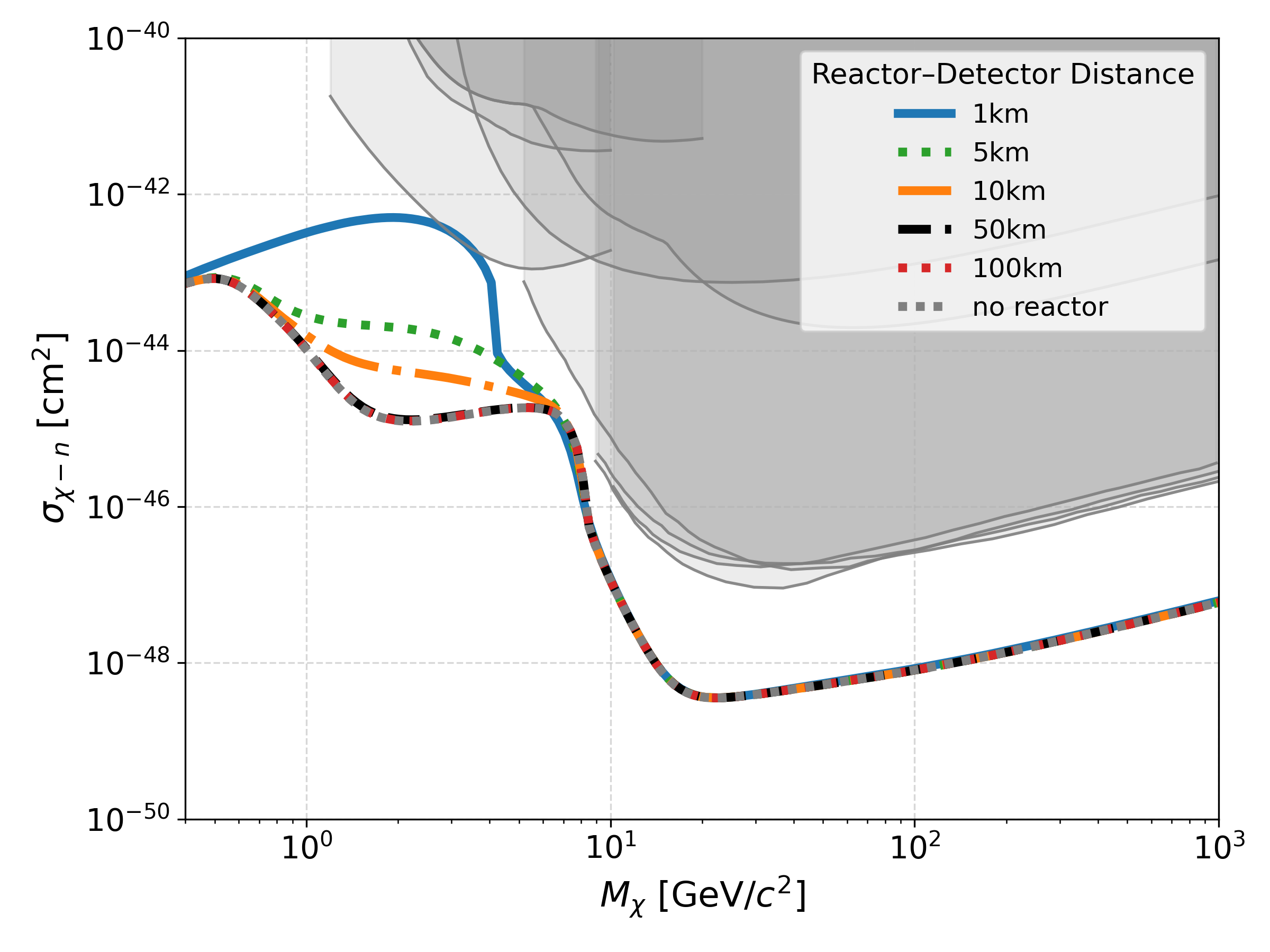}
    \includegraphics[width=0.48\linewidth]{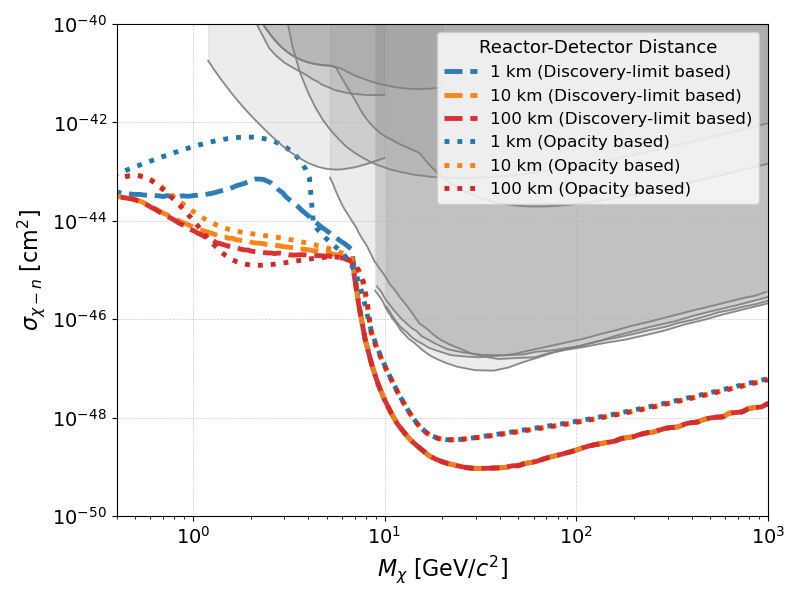}
    \caption{
    Effect of reactor proximity on low-mass dark matter searches. 
    \textbf{Upper Left:} discovery-limit–based neutrino floor. 
    \textbf{Upper Right:} opacity-based neutrino floor.  
\textbf{Bottom:} Comparison of discover-limit based and Opacity based neutrino floor at 1~km, 10~km and 100~km distances. The opacity-based method shows a stronger sensitivity to variations in reactor antineutrino flux, with a rapid degradation in sensitivity observed under both definitions for reactor–detector distances below 10 km.
    Grey regions indicate the parameter space excluded by leading direct-detection experiments, including SuperCDMS~\cite{CDMSliteR2_WIMP,CDMSliteR3_WIMP_PLR_2019}, EDELWEISS~\cite{EDELWEISS_WIMP_2019}, PICO~\cite{PICO_WIMP_2019}, DarkSide~\cite{DarkSide_2023_PRD_SI}, CRESST~\cite{CRESST_2019_WIMP,CRESST_2024_WIMP}, LZ~\cite{LZ_2023_WIMP}, PandaX~\cite{PandaX_2025_WIMP}, and XENON~\cite{XENONnT_2025_WIMP}. 
    The region between current experimental limits and the neutrino floor remains open and unexplored, representing the discovery window for future WIMP searches.  
    Short reactor–detector baselines significantly raise the neutrino floor for sub-10~GeV/$c^2$ WIMPs, narrowing the accessible parameter space and highlighting the importance of site selection in future low-mass dark matter experiments.
    }
    \label{fig:discovery_limit_based_vs_opacity_based}
\end{figure*}
Future direct-detection experiments targeting low-mass WIMPs ($M_\chi \lesssim 10$~GeV/$c^2$) must account for CE$\nu$NS backgrounds arising from nearby nuclear reactors.  
Reactor antineutrinos induce low-energy nuclear recoils that overlap with the expected WIMP signal region, effectively elevating the neutrino floor and reducing both sensitivity and discovery potential.

Figure~\ref{fig:discovery_limit_based_vs_opacity_based} shows the reactor-induced modification of the neutrino floor under the discovery-limit and opacity-based definitions.  
Both approaches exhibit similar qualitative trends but differ by up to an order of magnitude due to their distinct conceptual treatments~\cite{OHare:2021utq}.  

At short reactor–detector distances, the enhanced antineutrino flux dominates the low-energy recoil spectrum, substantially elevating the neutrino floor at low WIMP masses. As summarized in Table~\ref{tab:neutrino_floor_combined} for $M_\chi = 2.6$~GeV/$c^2$, the neutrino floor could increase by few orders of magnitude at baselines below $\sim$10~km and rapidly decreases at larger distances.
We also find that the opacity-based neutrino floor is more sensitive to variations in the antineutrino flux than the discovery-limit-based definition. To minimize the impact of reactor antineutrinos (limited to within a few percent), dark matter experiments probing sub-10~GeV/$c^2$ WIMPs should ideally be located more than 100~km away from a gigawatt-scale reactor.
These results emphasize the importance of accurate, site-specific background modeling in future sensitivity projections.  

Figure~\ref{fig:discovery_limit_based_vs_opacity_based} also presents the current exclusion limits from leading dark matter searches, where the shaded region denotes the parameter space already ruled out experimentally.  
The area between the present exclusion limits and the neutrino floor corresponds to the yet unexplored parameter space accessible to conventional direct detection experiments. Future searches aim to probe this region further. As evident, reactor proximity significantly reduces this discovery window in the sub-10~GeV/$c^2$ WIMP mass region, thereby constraining the potential for new discoveries in this mass range.

We also note that future experiments may be exposed to antineutrino fluxes originating from multi-reactor clusters, where individual reactors can have different thermal powers and baselines. In such environments, the total CE$\nu$NS background is simply the sum of the contributions from each reactor, weighted by their respective powers and source–detector distances. As discussed in Appendix~\ref{app:multiple_reactor}, a representative configuration consisting of reactors with 1.2~GW$_{\rm th}$ at 5~km, 2.4~GW$_{\rm th}$ at 10~km, and 3.6~GW$_{\rm th}$ at 50~km yields a combined antineutrino spectrum that is effectively equivalent to that from a single-reactor configuration with an effective baseline of $\sim$7~km and a total thermal power of 3.6~GW$_\mathrm{th}$. Consequently, the resulting neutrino floor is nearly indistinguishable from that of this equivalent single-reactor model.
Similarly, annual duty-cycle variations in reactor operation can be interpreted as a reduction in the effective thermal power, leading to a proportionally reduced antineutrino flux.
Finally, variations in the reactor antineutrino spectrum arising from changes in reactor fuel fission fractions have only a minor impact. As discussed in Appendix~\ref{app:neutrino_spectrum_shape_variation}, varying the fission composition modifies the predicted neutrino floor by at most $\sim$10\% at a reactor to detector distance of 5~km, far smaller than the order-of-magnitude enhancement of the neutrino floor relative to the no-reactor scenario.

\begin{table*}
\centering
\caption{
Comparison of neutrino floor values at $M_\chi = 2.6$~GeV/$c^2$ for different reactor–detector distances.
Both discovery-limit–based and opacity–based results are shown, along with scaling relative to the baseline (no-reactor) configuration and percentage deviations with respect to the baseline value.
Reactor antineutrinos can elevate the baseline neutrino floor by up to two orders of magnitude in the low-mass WIMP range for a gigawatt-scale source located approximately 1~km from the detector. 
}
\label{tab:neutrino_floor_combined}
\begin{ruledtabular}
\begin{tabular}{l@{\hspace{0.6cm}}ccc@{\hspace{0.8cm}}ccc}
\textrm{Reactor--detector distance} &
\multicolumn{3}{c}{\textrm{Discovery-limit--based}} &
\multicolumn{3}{c}{\textrm{Opacity--based}} \\
& \textrm{$\sigma_{\chi\text{--}n}$ [cm$^2$]} 
& \textrm{$\sigma/\sigma_{\text{baseline}}$} 
& \textrm{$\Delta$ [\%]} 
& \textrm{$\sigma_{\chi\text{--}n}$ [cm$^2$]} 
& \textrm{$\sigma/\sigma_{\text{baseline}}$} 
& \textrm{$\Delta$ [\%]} \\
[3pt]
\colrule
1 km        & $5.57\times10^{-44}$ & $25.79$ & $2479$ & $4.23\times10^{-43}$ & $326.15$ & $32515$ \\
5 km        & $6.28\times10^{-45}$ & $2.91$  & $191$  & $1.66\times10^{-44}$ & $12.77$  & $1177$  \\
10 km       & $3.32\times10^{-45}$ & $1.54$  & $54$   & $4.78\times10^{-45}$ & $3.68$   & $268$   \\
50 km       & $2.27\times10^{-45}$ & $1.05$  & $5$    & $1.34\times10^{-45}$ & $1.03$   & $3$     \\
100 km      & $2.20\times10^{-45}$ & $1.02$  & $2$    & $1.30\times10^{-45}$ & $1.00$   & $0$     \\
no reactor (baseline floor) & $2.16\times10^{-45}$ & $1.00$ & $-$ & $1.30\times10^{-45}$ & $1.00$ & $-$ \\
\end{tabular}
\end{ruledtabular}
\end{table*}

Although the impact of reactor antineutrino would be negligible for deep underground laboratories such as SNOLAB, where typical reactor baselines exceed $\sim$100~km~\cite{reactor_anti_neutrino_contribution_PRD_2024}, it becomes a vital consideration for next-generation low-threshold experiments and site planning.  
Accurate inclusion of reactor antineutrino fluxes in sensitivity projections is therefore essential for realistic assessments of discovery reach in the sub-10~GeV/$c^2$ WIMP mass regime.  
Mitigation strategies, including directional detection~\cite{Directional_search_OHare:2015utx,Directional_search_Grothaus:2014hja}, could further help preserve sensitivity in reactor-proximate environments.

\section{Summary and Outlook \label{Conclusion}}

We have investigated the impact of reactor antineutrinos on the sensitivity of future low-mass dark matter searches, primarily focusing on WIMPs with masses below 10~GeV/$c^2$.  
A SuperCDMS-like high-voltage germanium detector configuration was adopted, incorporating realistic detector response via quenching corrections using Lindhard model.  
Two widely adopted formulations of the neutrino floor-the discovery-limit and opacity-based approaches—were evaluated under consistent assumptions to quantify the effect of reactor proximity. This study represents the first comprehensive study of the neutrino floor that explicitly incorporates site-specific reactor proximity under both statistical definitions.
We find that proximity to gigawatt-scale reactors can significantly elevate the neutrino floor, particularly within $\sim$10~km, where enhancements of a few orders of magnitude arise in the sub-10~GeV/$c^2$ WIMP mass range.  
Such elevated backgrounds from reactor antineutrinos markedly reduce the discovery reach of low-mass dark matter experiments.  
To suppress this effect to the few-percent level, experiments aiming to probe sub-10~GeV/$c^2$ WIMPs should ideally be located at least $\sim$100~km from major reactors, beyond which the contribution becomes negligible.

Studies incorporating reasonable variations in the detector's quenching response, geoneutrino contribution, and uncertainties in the reactor antineutrino flux show that the dominant factor driving the loss of discovery potential is the geometric $1/L^{2}$ scaling of the reactor antineutrino flux.

These findings establish reactor proximity as an important design consideration for next-generation dark matter detectors.  
Future experiments exploring the low-mass WIMP parameter space must incorporate realistic, site-dependent reactor flux modeling when projecting sensitivities.  
Additionally, mitigation strategies such as directional detection could prove vital for maintaining sensitivity and fully exploiting the discovery potential of future low-threshold searches.

Recent work has shown that neutrino-induced Migdal events can generate an additional ``neutrino fog'' at sub-GeV/c$^2$ masses, where both CE$\nu$NS and WIMP interactions may trigger Migdal electron emission or excitation~\cite{migdal_neutrino_floor}. Such effects are especially relevant for inelastic dark-matter search channels~\cite{SuperCDMS_migdal_Brem_WIMP_2023}. In the present study, however, we restrict ourselves to the standard formulation of the neutrino floor based solely on elastic nuclear recoils, consistent with the original definitions in Refs.~\cite{Billard:2013qya,OHare:2021utq} and widely adopted in the literature. A dedicated investigation of how reactor-dependent backgrounds modify the Migdal-induced neutrino fog would require a distinct treatment of both the WIMP signal and CE$\nu$NS background models, and is beyond the scope of this work. This represents a natural direction for future study.

\section*{Acknowledgments}
We acknowledge the support of the Department of Atomic Energy (DAE) and the Department of Science and Technology (DST), Government of India. BM would like to also acknowledge the support received from the JC Bose Fellowship of the Anusandhan National Research Foundation (ANRF), Government of India. We also acknowledge the use of the Garuda and Kannad HPC cluster facilities at School of Physical Sciences, NISER.

\appendix

\section{Impact of Geoneutrino Flux Variations on the Neutrino Floor}
\label{app:geoneutrino_variation}

We adopt the geoneutrino flux from the China Jinping Underground Laboratory (CJPL) as a reference, which is relatively high due to the surrounding crustal composition. In comparison, SNOLAB, planned for next-generation SuperCDMS experiments, has a lower crustal geoneutrino flux, approximately half that of CJPL, reflecting its distinct local crustal and geological environment~\cite{geo_neutrino_SNOLAB_vs_CJPL}. 

\begin{figure}[h]
\centering
\includegraphics[width=0.95\linewidth]{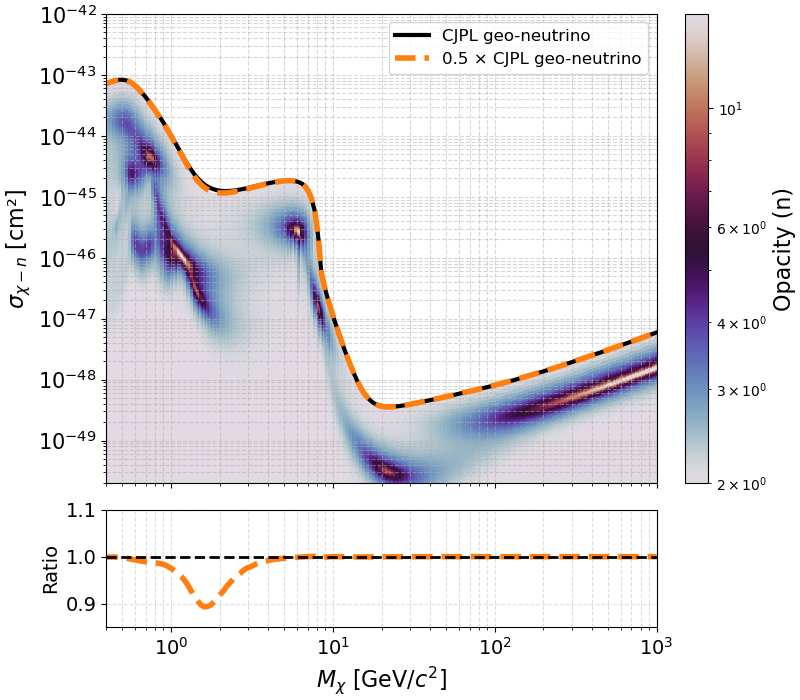}
\caption{Dependence of a factor-of-two variation in the geoneutrino flux on the neutrino floor in the absence of reactor contributions. The variation produces up to a $\sim$10\% change near $M_\chi \sim 1.7~\mathrm{GeV}/c^2$.}
\label{fig:geo_neutrino_variation}
\end{figure}

To quantify the impact of site-dependent geoneutrino flux uncertainties, we performed a variation study by scaling the geoneutrino contribution by a factor of two, in the absence of reactor neutrino contributions. As shown in Fig.~\ref{fig:geo_neutrino_variation}, this variation modifies the neutrino floor by up to $\sim$10\% near $M_\chi \sim 1.7~\mathrm{GeV}/c^2$. Overall, the geoneutrino contribution remains subdominant compared to the solar neutrino background.

For detectors exposed to a gigawatt-scale reactor at $\sim$10~km, the geoneutrino effect becomes subdominant, quickly suppressed by the reactor antineutrino flux. This demonstrates that while local crustal composition can slightly affect the neutrino floor, reactor proximity dominates for near-reactor sites.

\section{Effect of Quenching on the Neutrino Floor}
\label{app:quenching_effect}

To evaluate the effect of quenching-model uncertainties on the neutrino floor, we studied the no-reactor scenario as a representative benchmark. The analysis uses the Lindhard parametrization of the ionization yield, with the quenching parameter reported by CONUS+, \(k = 0.162 \pm 0.004\)~\cite{COUNSplus_CEvNS}. To assess the sensitivity of the neutrino floor to this uncertainty, we varied the quenching parameter within its \(3\sigma\) range. This variation modifies both the magnitude and the shape of the corresponding Lindhard quenching curve.

\begin{figure}[h!]
    \centering
    \includegraphics[width=\linewidth]{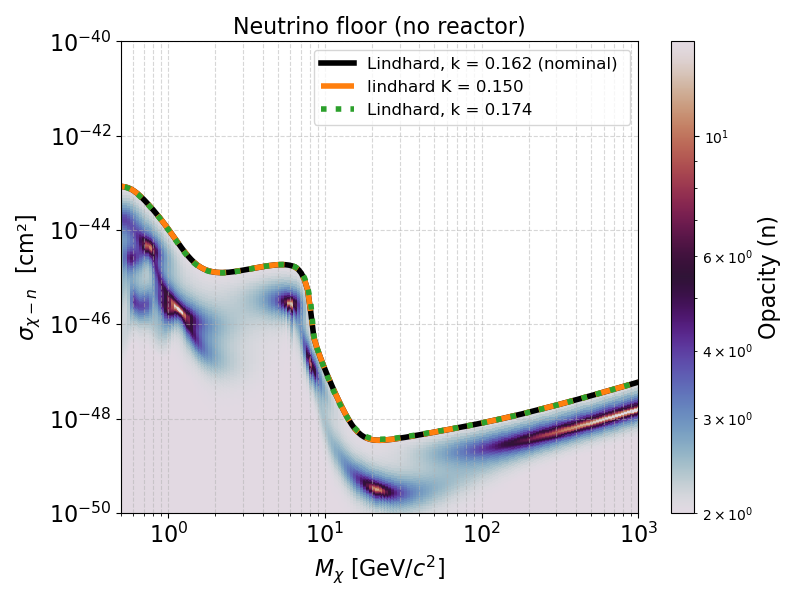}
    \caption{Impact of varying the Lindhard quenching parameter \(k\) within its \(3\sigma\) range on the neutrino floor in the no-reactor scenario. The comparison includes the nominal CONUS+ value \(k=0.162\) and the shifted values \(k{\,+3\sigma}=0.174\) and \(k{\,-3\sigma}=0.150\). The resulting neutrino floors are nearly identical, indicating the negligible effect of quenching uncertainties.}
    \label{fig:quenching_variation}
\end{figure}

Figure~\ref{fig:quenching_variation} shows the resulting neutrino floors obtained for the nominal, \(k + 3\sigma\), and \(k - 3\sigma\) cases. The resulting neutrino floors are consistent across the full WIMP-mass range, demonstrating that the neutrino floor is robust against reasonable variations in the quenching parameter. This behavior is consistent with the expectation that both CE$\nu$NS and WIMP-induced spectra are modified in a similar manner by this detector effect.

Although the quenching uncertainty has a negligible impact on the neutrino-floor calculation itself, a realistic quenching model remains essential for predicting measurable recoil spectra and for performing spectral analyses with experimental data.

\section{Validation of reactor CE$\nu$NS differential rate}
\label{app:reactor_CEnuNS_validation}

We validated our CE$\nu$NS differential event rate modeling for the KKL reactor (Switzerland) using the experimentally observed reactor antineutrino spectrum from the CONUS+ experiment.
CONUS+ is a dedicated CE$\nu$NS search experiment employing kg-scale ultra-low-threshold (160-180 eV$_{ee}$) HPGe detectors to measure neutrino–nucleus interactions. The setup is located approximately 20.7~m from the KKL reactor core and has reported the first observation of reactor antineutrino-induced CE$\nu$NS signals with a significance of 3.7$\sigma$~\cite{CONUS+_first_observation_2025}.

To cross-check our reactor CE$\nu$NS model, we computed the expected differential event rate for a germanium detector placed at 20.7~m from the KKL core and compared it with the CONUS+ measured spectrum.
Since the HPGe detectors in CONUS+ measure ionization signals, the nuclear recoil energy from CE$\nu$NS interactions undergoes quenching. Accordingly, we converted the predicted nuclear recoil spectrum into the corresponding electron-equivalent energy using the Lindhard model with $k = 0.162$.

\begin{figure}[h!]
\centering
\includegraphics[width=1\linewidth]{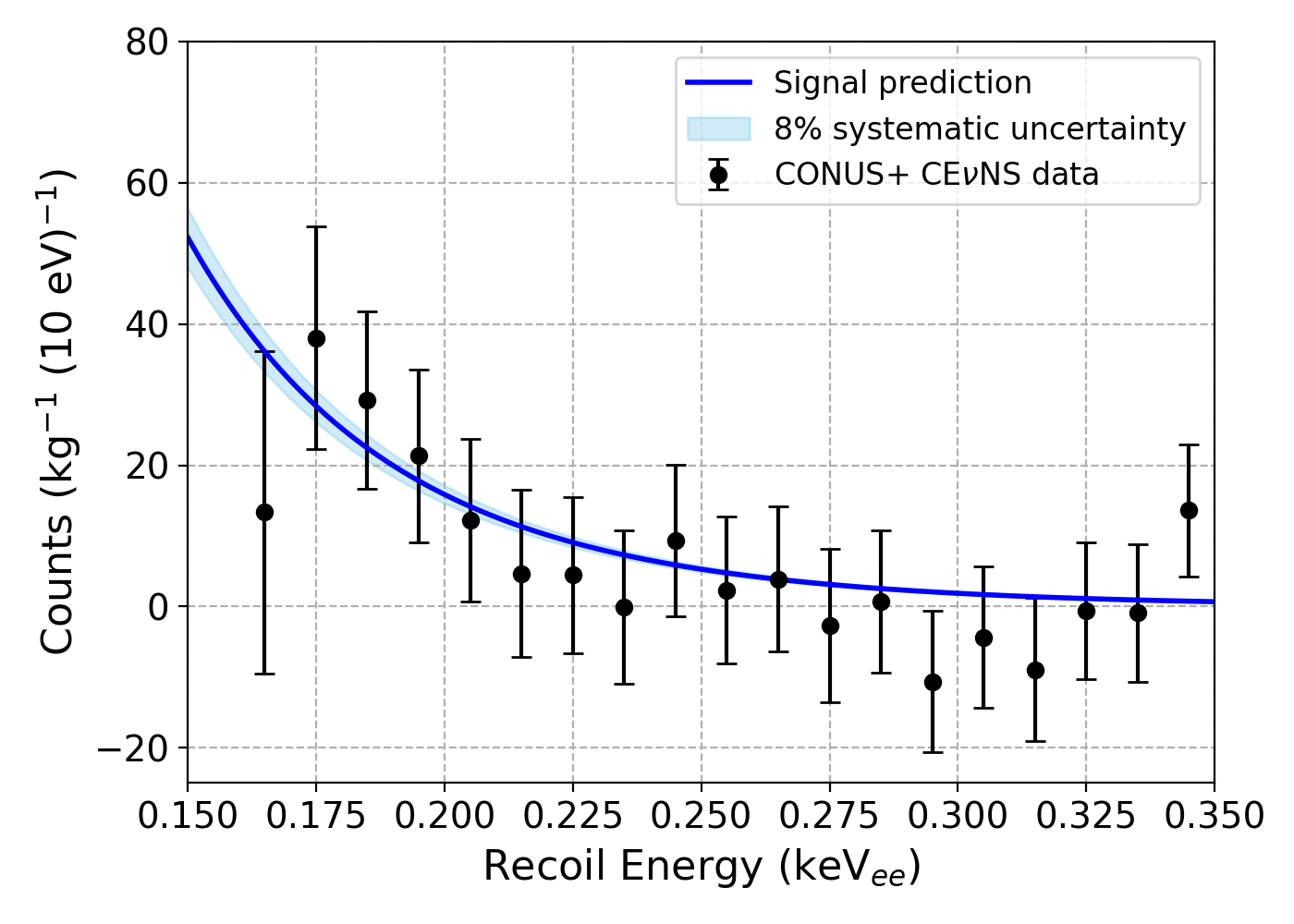}
\caption{Comparison between the predicted and CONUS+ measured CE$\nu$NS differential spectra for the KKL reactor at a baseline of 20.7~m. 
The prediction includes quenching corrections using the Lindhard model with $k=0.162$, and the shaded band represents an 8\% systematic uncertainty in the modeled rate.
}
\label{fig:reactor_anti-neutrino_spectra_comparison}
\end{figure}

As shown in Fig.~\ref{fig:reactor_anti-neutrino_spectra_comparison}, our predicted differential event rate exhibits good agreement with the CONUS+ experimental results, thereby validating both our reactor antineutrino flux model and the CE$\nu$NS differential spectrum modeling in terms of electron-equivalent measurable energy.

\section{Comparison of baseline neutrino floor}
\label{app:baseline_comparison}

\begin{figure}[h!]
    \centering
    \includegraphics[width=\linewidth]{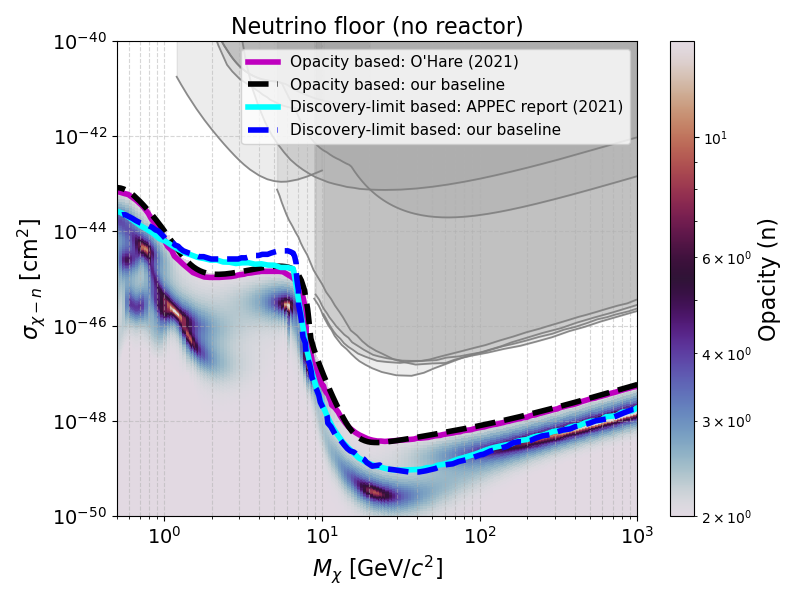}
    \caption{Comparison of the baseline neutrino floor (no-reactor case) with the published discovery-limit--based~\cite{Billard:APPEC_report} and opacity-based~\cite{OHare:2021utq} results for a Ge detector. Overall, our results reproduce the published trends. The lower neutrino floor in the discovery-limit case near a WIMP mass of $6~\mathrm{GeV}/c^{2}$ arises from the much smaller (2\%) $^8$B flux systematic uncertainty used in this work compared to the 16\% adopted in Ref.~\cite{Billard:APPEC_report}.  Grey regions indicate the parameter space excluded by leading direct-detection experiments, including SuperCDMS~\cite{CDMSliteR2_WIMP,CDMSliteR3_WIMP_PLR_2019}, EDELWEISS~\cite{EDELWEISS_WIMP_2019}, PICO~\cite{PICO_WIMP_2019}, DarkSide~\cite{DarkSide_2023_PRD_SI}, CRESST~\cite{CRESST_2019_WIMP,CRESST_2024_WIMP}, LZ~\cite{LZ_2023_WIMP}, PandaX~\cite{PandaX_2025_WIMP}, and XENON~\cite{XENONnT_2025_WIMP}.}
    \label{fig:baseline_limit_comparison}
\end{figure}

We compared our baseline neutrino floor results---i.e., the neutrino floor without the effect of reactor antineutrinos---with previously published results for a Ge detector. The comparison was performed for both the discovery-limit--based~\cite{Billard:APPEC_report} and opacity-based~\cite{OHare:2021utq} neutrino floors and shown in Fig. \ref{fig:baseline_limit_comparison}.

For the discovery-limit case, our result is consistent with the APPEC Committee Report (2021)~\cite{Billard:APPEC_report}, except near a WIMP mass of $\sim 6~\mathrm{GeV}/c^{2}$. This deviation originates from the significantly reduced systematic uncertainty used for the $^8$B solar-neutrino flux in our study (2\%, see Table~\ref{tab:flux_uncertainties}) compared to the 16\% used in the APPEC report. As discussed in Fig.~\ref{fig:reactor_systematics}, a reduction in systematic uncertainty by an order of magnitude can lower the neutrino floor, thereby expanding the discovery potential for direct dark matter detection. We have verified that adopting a 16\% uncertainty for the $^8$B flux raises the floor, reproducing the result reported by the APPEC community.  

For the opacity-based neutrino floor, our baseline result shows reasonable agreement with the published calculation of Ref.~\cite{OHare:2021utq} for a Ge detector. It is worth noting that the O’Hare calculation was performed for the Gran Sasso site, accounting for both geo- and reactor antineutrino contributions at that site. The reactor antineutrino component was found to be negligible owing to the large distances to nearby reactors (\(\mathcal{O}(100~\mathrm{km})\)). This is consistent with our findings, which indicate that such distant reactors have an insignificant effect on the neutrino floor.

\section{Influence of Multiple Nearby Reactors on the Neutrino Floor}
\label{app:multiple_reactor}

Experiments often encounter antineutrino fluxes originating from multiple reactors with different thermal powers and baselines. To evaluate the implications of such multi-reactor environments for our study, we computed the neutrino floor for a representative mixed-reactor configuration in which the total antineutrino flux at the detector is obtained by summing the contributions from all reactors. Each contribution is weighted by its thermal power and by the inverse square of the source-to-detector distance, producing a combined CE$\nu$NS spectrum that reflects the full cluster.

\begin{figure}[h!]
    \centering
    \includegraphics[width=\linewidth]{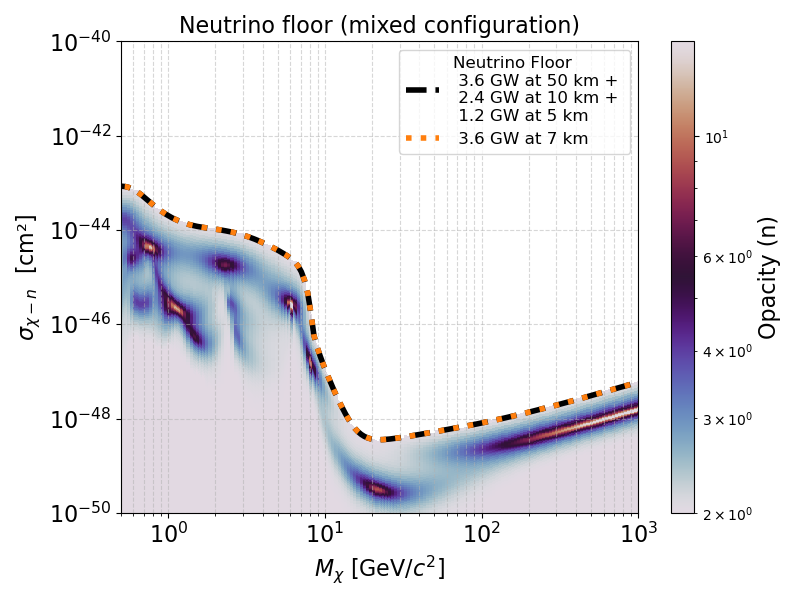}
    \caption{Neutrino-floor prediction for the mixed-reactor scenario consisting of three reactors at different baselines and thermal powers (1.2~GW$_\text{th}$ at 5~km, 2.4~GW$_\text{th}$ at 10~km, and 3.6~GW$_\text{th}$ at 50~km). The combined CE$\nu$NS spectrum from this reactor cluster is effectively equivalent to that produced by a single 3.6~GW$_\text{th}$ reactor located at an effective distance of $\sim$7~km, resulting in an indistinguishable neutrino floor.}
    \label{fig:mixed_baseline_config}
\end{figure}

The specific configuration considered consists of a 1.2~GW$_\text{th}$ reactor at 5~km, a 2.4~GW$_\text{th}$ reactor at 10~km, and a 3.6~GW$_\text{th}$ reactor at 50~km. Figure~\ref{fig:mixed_baseline_config} shows the neutrino floor obtained for this mixed configuration. The resulting total antineutrino spectrum is equivalent to that produced by a single 3.6~GW$_\text{th}$ reactor located at an effective distance of approximately 7~km. Consequently, the neutrino floor obtained for the mixed-reactor case is effectively identical to that of the corresponding single-reactor effective configuration.

This demonstrates that, for the purposes of neutrino-floor estimation, the mixed-reactor configuration considered here is effectively equivalent to a single-reactor scenario with comparable flux normalization. Similarly, annual duty-cycle averaging of reactor operations can lead to a reduced effective antineutrino flux, which may be treated as a single-reactor configuration with a correspondingly lower thermal power.

\section{Impact of Reactor Antineutrino Spectrum Shape Variations}
\label{app:neutrino_spectrum_shape_variation}

Nearby reactors may operate with different fuel compositions, leading to variations in their fission fractions and, consequently, in the shape of the emitted antineutrino spectrum. As discussed in Sec.~\ref{subsec:reactor_antineutrino}, changes in the relative contributions of $^{235}$U, $^{238}$U, $^{239}$Pu, and $^{241}$Pu affect both the normalization and energy dependence of the flux. To quantify this effect, we consider two representative configurations. The first corresponds to a commercial light-water reactor with fission fractions ${^{235}\mathrm{U}: 0.53,; ^{238}\mathrm{U}: 0.08,; ^{239}\mathrm{Pu}: 0.32,; ^{241}\mathrm{Pu}: 0.07}$ \cite{conus+_experiment_2024}. The second represents a highly enriched research reactor such as HFIR, in which $^{235}$U accounts for approximately 99\% of all fissions~\cite{HFIR_fission_frac}.
\begin{figure}
\centering
\includegraphics[width=\linewidth]{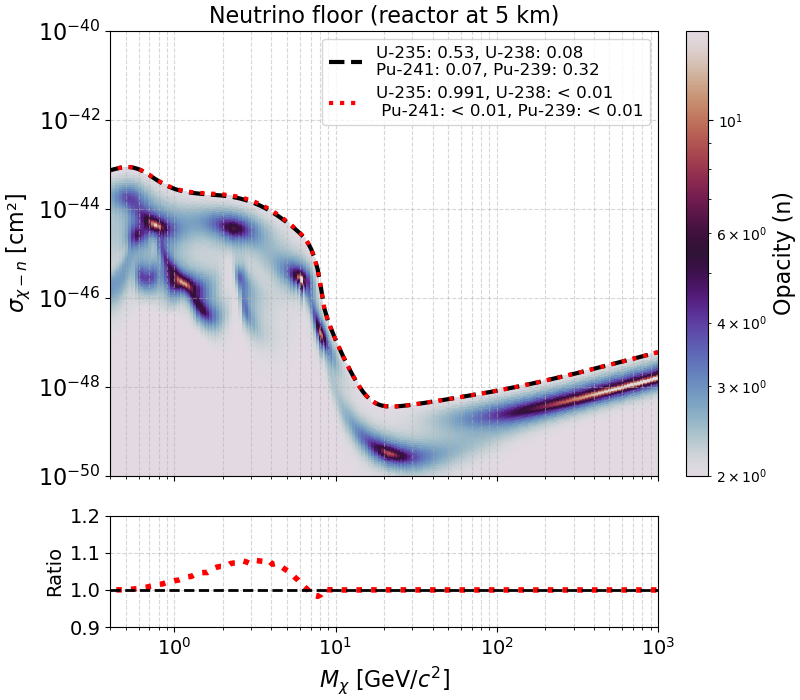}
\caption{Comparison of neutrino-floor predictions at a reactor–detector distance of 5~km for a commercial (KKL-like) fission-fraction configuration and a highly enriched $^{235}$U  (HFIR-like) research-reactor configuration. Both assume a thermal power of 3.6~GW$_\text{th}$.}
\label{fig:fission_frac_variation}
\end{figure}

For both scenarios, we compute the CE$\nu$NS event rate assuming a thermal power of 3.6~GW$_\text{th}$ and a reactor–detector separation of 5~km. The resulting variation in the neutrino floor is at the level of $\sim$10\%, which is well below the order-of-magnitude enhancement observed at 5~km relative to the no-reactor baseline. Therefore, uncertainties in reactor fuel composition do not constitute a dominant systematic uncertainty in the determination of the neutrino floor.

\bibliography{main}

@article{first_proposed_DM:1933gu,
    author = "Zwicky, F.",
    title = "{Die Rotverschiebung von extragalaktischen Nebeln}",
    doi = "10.1007/s10714-008-0707-4",
    journal = "Helv. Phys. Acta",
    volume = "6",
    pages = "110--127",
    year = "1933"
}

@article{rotation_curve_1970,
    author = "Rubin, Vera C. and Ford, Jr., W. Kent",
    title = "{Rotation of the Andromeda Nebula from a Spectroscopic Survey of Emission Regions}",
    doi = "10.1086/150317",
    journal = "Astrophys. J.",
    volume = "159",
    pages = "379--403",
    year = "1970"
}

@article{bullet_cluster_2006,
    author = "Clowe, Douglas and Bradac, Marusa and Gonzalez, Anthony H. and Markevitch, Maxim and Randall, Scott W. and Jones, Christine and Zaritsky, Dennis",
    title = "{A direct empirical proof of the existence of dark matter}",
    eprint = "astro-ph/0608407",
    archivePrefix = "arXiv",
    reportNumber = "SLAC-PUB-12078",
    doi = "10.1086/508162",
    journal = "Astrophys. J. Lett.",
    volume = "648",
    pages = "L109--L113",
    year = "2006"
}

@article{Planck_2015_CMB,
    author = "Ade, P. A. R. and others",
    collaboration = "Planck",
    title = "{Planck 2015 results. XIII. Cosmological parameters}",
    eprint = "1502.01589",
    archivePrefix = "arXiv",
    primaryClass = "astro-ph.CO",
    doi = "10.1051/0004-6361/201525830",
    journal = "Astron. Astrophys.",
    volume = "594",
    pages = "A13",
    year = "2016"
}

@article{WIMP_1_1984,
    author = "Steigman, Gary and Turner, Michael S.",
    title = "{Cosmological Constraints on the Properties of Weakly Interacting Massive Particles}",
    reportNumber = "FERMILAB-PUB-84-110-A, BA-84-33",
    doi = "10.1016/0550-3213(85)90537-1",
    journal = "Nucl. Phys. B",
    volume = "253",
    pages = "375--386",
    year = "1985"
}

@article{WIMP_2_1995_supersymmetric,
    author = "Jungman, Gerard and Kamionkowski, Marc and Griest, Kim",
    title = "{Supersymmetric dark matter}",
    eprint = "hep-ph/9506380",
    archivePrefix = "arXiv",
    reportNumber = "SU-4240-605, UCSD-PTH-95-02, IASSNS-HEP-95-14, CU-TP-677",
    doi = "10.1016/0370-1573(95)00058-5",
    journal = "Phys. Rept.",
    volume = "267",
    pages = "195--373",
    year = "1996"
}

@article{Relic_WIMP_abundances_2012,
    author = "Steigman, Gary and Dasgupta, Basudeb and Beacom, John F.",
    title = "{Precise Relic WIMP Abundance and its Impact on Searches for Dark Matter Annihilation}",
    eprint = "1204.3622",
    archivePrefix = "arXiv",
    primaryClass = "hep-ph",
    doi = "10.1103/PhysRevD.86.023506",
    journal = "Phys. Rev. D",
    volume = "86",
    pages = "023506",
    year = "2012"
}

@article{LZ_2023_WIMP,
    author = "Aalbers, J. and others",
    collaboration = "LZ",
    title = "{First Dark Matter Search Results from the LUX-ZEPLIN (LZ) Experiment}",
    eprint = "2207.03764",
    archivePrefix = "arXiv",
    primaryClass = "hep-ex",
    doi = "10.1103/PhysRevLett.131.041002",
    journal = "Phys. Rev. Lett.",
    volume = "131",
    number = "4",
    pages = "041002",
    year = "2023"
}

@article{PandaX_2025_WIMP,
    author = "Bo, Zihao and others",
    collaboration = "PandaX",
    title = "{Dark Matter Search Results from 1.54\,\,Tonne\textperiodcentered{}Year Exposure of PandaX-4T}",
    eprint = "2408.00664",
    archivePrefix = "arXiv",
    primaryClass = "hep-ex",
    doi = "10.1103/PhysRevLett.134.011805",
    journal = "Phys. Rev. Lett.",
    volume = "134",
    number = "1",
    pages = "011805",
    year = "2025"
}

@article{CRESST_2025_projected_sensitivity,
  title={The CRESST experiment: towards the next-generation of sub-GeV direct dark matter detection},
  author={Angloher, G and others},
  doi ="arXiv preprint arXiv:2505.01183",
  year={2025}
}

@article{SuperCDMS_SNOALB_projected_SNOWMASS_2021,
  author       = {M. F. Albakry and others},
  title        = {A Strategy for Low-Mass Dark Matter Searches with Cryogenic Detectors in the SuperCDMS SNOLAB Facility},
  journal      = {arXiv preprint arXiv:2203.08463},
  year         = {2022},
  url          = {https://arxiv.org/abs/2203.08463}
}

@article{xenon2024xenonnt_status,
  author       = {E. Aprile and others},
  collaboration= {XENON Collaboration},
  title        = {The XENONnT dark matter experiment},
  journal      = {Eur. Phys. J. C},
  volume       = {84},
  number       = {8},
  pages        = {784},
  year         = {2024},
  doi          = {10.1140/epjc/s10052-024-12982-5},
  eprint       = {2402.10446},
  archivePrefix= {arXiv},
  primaryClass = {physics.ins-det}
}

@article{freedman,
  title = {Coherent effects of a weak neutral current},
  author = {Freedman, Daniel Z.},
  journal = {Phys. Rev. D},
  volume = {9},
  issue = {5},
  pages = {1389--1392},
  numpages = {0},
  year = {1974},
  month = {Mar},
  publisher = {American Physical Society},
  doi = {10.1103/PhysRevD.9.1389},
  url = {https://link.aps.org/doi/10.1103/PhysRevD.9.1389}
}

@article{Billard:2013qya,
    author = "Billard, J. and Strigari, L. and Figueroa-Feliciano, E.",
    title = "{Implication of neutrino backgrounds on the reach of next generation dark matter direct detection experiments}",
    eprint = "1307.5458",
    archivePrefix = "arXiv",
    primaryClass = "hep-ph",
    doi = "10.1103/PhysRevD.89.023524",
    journal = "Phys. Rev. D",
    volume = "89",
    number = "2",
    pages = "023524",
    year = "2014"
}

@article{Ruppin_followup:2014bra,
    author = "Ruppin, F. and Billard, J. and Figueroa-Feliciano, E. and Strigari, L.",
    title = "{Complementarity of dark matter detectors in light of the neutrino background}",
    eprint = "1408.3581",
    archivePrefix = "arXiv",
    primaryClass = "hep-ph",
    doi = "10.1103/PhysRevD.90.083510",
    journal = "Phys. Rev. D",
    volume = "90",
    number = "8",
    pages = "083510",
    year = "2014"
}

@article{OHare:2021utq,
    author = "O'Hare, Ciaran A. J.",
    title = "{New Definition of the Neutrino Floor for Direct Dark Matter Searches}",
    eprint = "2109.03116",
    archivePrefix = "arXiv",
    primaryClass = "hep-ph",
    doi = "10.1103/PhysRevLett.127.251802",
    journal = "Phys. Rev. Lett.",
    volume = "127",
    number = "25",
    pages = "251802",
    year = "2021"
}

@article{CJPL_neutrino_floor_Fan:2025sde,
    author = "Fan, Yingjie and Liu, Xuewen and Zhou, Ning",
    title = "{Characterizing the CJPL{\textquoteright}s site-specific neutrino floor as the neutrino fog boundary*}",
    eprint = "2503.22155",
    archivePrefix = "arXiv",
    primaryClass = "hep-ph",
    doi = "10.1088/1674-1137/ade65e",
    journal = "Chin. Phys.",
    volume = "49",
    number = "10",
    pages = "103001",
    year = "2025"
}

@article{Overcome_neutrino_floor:2020lva,
    author = "O'Hare, Ciaran A. J.",
    title = "{Can we overcome the neutrino floor at high masses?}",
    eprint = "2002.07499",
    archivePrefix = "arXiv",
    primaryClass = "astro-ph.CO",
    reportNumber = "CPPC-2020-13",
    doi = "10.1103/PhysRevD.102.063024",
    journal = "Phys. Rev. D",
    volume = "102",
    number = "6",
    pages = "063024",
    year = "2020"
}

@article{COHERENT_CEvNS,
author = {D. Akimov  et.al  and COHERENT Collaboration},
title = {Observation of coherent elastic neutrino-nucleus scattering},
journal = {Science},
volume = {357},
number = {6356},
pages = {1123-1126},
year = {2017},
doi = {10.1126/science.aao0990},
URL = {https://www.science.org/doi/abs/10.1126/science.aao0990},
eprint = {https://www.science.org/doi/pdf/10.1126/science.aao0990}
}

@article{COUNSplus_CEvNS,
    author = "Ackermann, N. and others",
    title = "{Direct observation of coherent elastic antineutrino-nucleus scattering}",
    eprint = "2501.05206",
    archivePrefix = "arXiv",
    primaryClass = "hep-ex",
    doi = "10.1038/s41586-025-09322-2",
    journal = "Nature",
    volume = "643",
    pages = "1229--1233",
    year = "2025"
}

@article{WIMP_rate_Lewin,
  author        = {Lewin, J. D. and Smith, P. F.},
  title         = {Review of mathematics, numerical factors, and corrections for dark matter experiments based on elastic nuclear recoil},
  journal       = {Astropart. Phys.},
  volume        = {6},
  pages         = {87--112},
  year          = {1996},
  doi           = {10.1016/S0927-6505(96)00047-3}
}

@article{Standard_halo_model,
  author        = {Lewin, J. D. and Smith, P. F.},
  title         = {Review of mathematics, numerical factors, and corrections for dark matter experiments based on elastic nuclear recoil},
  journal       = {Astropart. Phys.},
  volume        = {6},
  pages         = {87--112},
  year          = {1996},
  doi           = {10.1016/S0927-6505(96)00047-3}
}

@book{lindhard_quenching_1963,
  title={Range concepts and heavy ion ranges},
  author={Lindhard, Jens and Scharff, Morten and Schi{\o}tt, Hans E},
  volume={33},
  year={1963},
  publisher={Munksgaard Copenhagen}
}

@article{CEvNS_Crosssection_Scholberg:2005qs,
    author = "Scholberg, Kate",
    title = "{Prospects for measuring coherent neutrino-nucleus elastic scattering at a stopped-pion neutrino source}",
    eprint = "hep-ex/0511042",
    archivePrefix = "arXiv",
    doi = "10.1103/PhysRevD.73.033005",
    journal = "Phys. Rev. D",
    volume = "73",
    pages = "033005",
    year = "2006"
}

@article{Geo_neutrino_Gelmini:2018gqa,
    author = "Gelmini, Graciela B. and Takhistov, Volodymyr and Witte, Samuel J.",
    title = "{Geoneutrinos in Large Direct Detection Experiments}",
    eprint = "1812.05550",
    archivePrefix = "arXiv",
    primaryClass = "hep-ph",
    doi = "10.1103/PhysRevD.99.093009",
    journal = "Phys. Rev. D",
    volume = "99",
    number = "9",
    pages = "093009",
    year = "2019"
}

@article{Hubber_reactor_spectrum_PhysRevC.84.024617,
  title = {Determination of antineutrino spectra from nuclear reactors},
  author = {Huber, Patrick},
  journal = {Phys. Rev. C},
  volume = {84},
  issue = {2},
  pages = {024617},
  numpages = {16},
  year = {2011},
  month = {Aug},
  publisher = {American Physical Society},
  doi = {10.1103/PhysRevC.84.024617},
  url = {https://link.aps.org/doi/10.1103/PhysRevC.84.024617}
}

@article{CONNIE_anti_neutrino_spectrum:2019xid,
    author = "Aguilar-Arevalo, Alexis and others",
    collaboration = "CONNIE",
    title = "{Search for light mediators in the low-energy data of the CONNIE reactor neutrino experiment}",
    eprint = "1910.04951",
    archivePrefix = "arXiv",
    primaryClass = "hep-ex",
    reportNumber = "FERMILAB-PUB-19-525-AE-PPD-SCD",
    doi = "10.1007/JHEP04(2020)054",
    journal = "JHEP",
    volume = "04",
    pages = "054",
    year = "2020"
}

@article{TEXONO_U238_capture:2006xds,
    author = "Wong, H. T. and others",
    collaboration = "TEXONO",
    title = "{A Search of Neutrino Magnetic Moments with a High-Purity Germanium Detector at the Kuo-Sheng Nuclear Power Station}",
    eprint = "hep-ex/0605006",
    archivePrefix = "arXiv",
    reportNumber = "AS-TEXONO-06-03",
    doi = "10.1103/PhysRevD.75.012001",
    journal = "Phys. Rev. D",
    volume = "75",
    pages = "012001",
    year = "2007"
}

@article{conus+_experiment_2024,
  title={CONUS+ experiment},
  author={Ackermann, N and Armbruster, S and Bonet, H and Buck, C and F{\"u}lber, K and Hakenm{\"u}ller, J and Hempfling, J and Heusser, G and Lindner, M and Maneschg, W and others},
  journal={The European Physical Journal C},
  volume={84},
  number={12},
  pages={1265},
  year={2024},
  publisher={Springer}
}

@article{cowan2011asymptotic,
  title={Asymptotic formulae for likelihood-based tests of new physics},
  author={Cowan, Glen and Cranmer, Kyle and Gross, Eilam and Vitells, Ofer},
  journal={The European Physical Journal C},
  volume={71},
  number={2},
  pages={1554},
  year={2011},
  publisher={Springer}
}

@article{Directional_search_OHare:2015utx,
    author = "O'Hare, Ciaran A. J. and Green, Anne M. and Billard, Julien and Figueroa-Feliciano, Enectali and Strigari, Louis E.",
    title = "{Readout strategies for directional dark matter detection beyond the neutrino background}",
    eprint = "1505.08061",
    archivePrefix = "arXiv",
    primaryClass = "astro-ph.CO",
    doi = "10.1103/PhysRevD.92.063518",
    journal = "Phys. Rev. D",
    volume = "92",
    number = "6",
    pages = "063518",
    year = "2015"
}

@article{Directional_search_Grothaus:2014hja,
    author = "Grothaus, Philipp and Fairbairn, Malcolm and Monroe, Jocelyn",
    title = "{Directional Dark Matter Detection Beyond the Neutrino Bound}",
    eprint = "1406.5047",
    archivePrefix = "arXiv",
    primaryClass = "hep-ph",
    reportNumber = "KCL-PH-TH-2014-28, LCTS-2014-27",
    doi = "10.1103/PhysRevD.90.055018",
    journal = "Phys. Rev. D",
    volume = "90",
    number = "5",
    pages = "055018",
    year = "2014"
}

@article{reactor_anti_neutrino_contribution_PRD_2024,
    author = "Aristizabal Sierra, D. and De Romeri, Valentina and Ternes, Christoph A.",
    title = "{Reactor neutrino background in next-generation dark matter detectors}",
    eprint = "2402.06416",
    archivePrefix = "arXiv",
    primaryClass = "hep-ph",
    doi = "10.1103/PhysRevD.109.115026",
    journal = "Phys. Rev. D",
    volume = "109",
    number = "11",
    pages = "115026",
    year = "2024"
}

@article{CONUS_reactor_anti-neutrino_observation_Ackermann:2025obx,
    author = "Ackermann, N. and others",
    title = "{Direct observation of coherent elastic antineutrino{\textendash}nucleus scattering}",
    eprint = "2501.05206",
    archivePrefix = "arXiv",
    primaryClass = "hep-ex",
    doi = "10.1038/s41586-025-09322-2",
    journal = "Nature",
    volume = "643",
    number = "8074",
    pages = "1229--1233",
    year = "2025"
}

@article{Billard:APPEC_report,
    author = "Billard, Julien and others",
    title = "{Direct detection of dark matter{\textemdash}APPEC committee report*}",
    eprint = "2104.07634",
    archivePrefix = "arXiv",
    primaryClass = "hep-ex",
    doi = "10.1088/1361-6633/ac5754",
    journal = "Rept. Prog. Phys.",
    volume = "85",
    number = "5",
    pages = "056201",
    year = "2022"
}

@article{SuperCDMS:HVeV_compton_steps,
  title = {Low-energy calibration of SuperCDMS high-voltage eV-resolution cryogenic silicon calorimeters using Compton steps},
  author = {al., M. F. Albakry et},
  journal = {Phys. Rev. D},
  pages = {--},
  year = {2025},
  month = {Oct},
  publisher = {American Physical Society},
  doi = {10.1103/jj7w-gkgg},
  url = {https://link.aps.org/doi/10.1103/jj7w-gkgg}
}

@article{HVeVR1,
  title={{First dark matter constraints from a SuperCDMS single-charge sensitive detector}},
  author={Agnese, R and others},
  collaboration = {SuperCDMS Collaboration},
  journal={Physical Review Letters},
  volume={121},
  number={5},
  pages={051301},
  year={2018},
  doi = "10.1103/PhysRevLett.121.051301",
  publisher={APS}
}

@article{SuperCDMS:HVeV4_DM,
    author = "Albakry, M. F. and others",
    collaboration = "SuperCDMS",
    title = "{Search for low-mass electron-recoil dark matter using a single-charge sensitive SuperCDMS-HVeV Detector}",
    eprint = "2509.03608",
    archivePrefix = "arXiv",
    primaryClass = "hep-ex",
    reportNumber = "FERMILAB-PUB-25-0643-PPD",
    month = "9",
    year = "2025"
}

@article{CRESST_2024_WIMP,
    author = "Angloher, G. and others",
    collaboration = "CRESST",
    title = "{First observation of single photons in a CRESST detector and new dark matter exclusion limits}",
    eprint = "2405.06527",
    archivePrefix = "arXiv",
    primaryClass = "astro-ph.CO",
    doi = "10.1103/PhysRevD.110.083038",
    journal = "Phys. Rev. D",
    volume = "110",
    number = "8",
    pages = "083038",
    year = "2024"
}

@article{CDMSliteR2_WIMP,
    author = "Agnese, R. and others",
    collaboration = "SuperCDMS",
    title = "{New Results from the Search for Low-Mass Weakly Interacting Massive Particles with the CDMS Low Ionization Threshold Experiment}",
    eprint = "1509.02448",
    archivePrefix = "arXiv",
    primaryClass = "astro-ph.CO",
    reportNumber = "IPPP-15-56, DCTP-15-112, FERMILAB-PUB-15-394-AE",
    doi = "10.1103/PhysRevLett.116.071301",
    journal = "Phys. Rev. Lett.",
    volume = "116",
    number = "7",
    pages = "071301",
    year = "2016"
}

@article{CDMSliteR3_WIMP_PLR_2019,
    author = "Agnese, R. and others",
    collaboration = "SuperCDMS",
    title = "{Search for Low-Mass Dark Matter with CDMSlite Using a Profile Likelihood Fit}",
    eprint = "1808.09098",
    archivePrefix = "arXiv",
    primaryClass = "astro-ph.CO",
    reportNumber = "FERMILAB-PUB-18-435-AE",
    doi = "10.1103/PhysRevD.99.062001",
    journal = "Phys. Rev. D",
    volume = "99",
    number = "6",
    pages = "062001",
    year = "2019"
}

@article{SuperCDMS_migdal_Brem_WIMP_2023,
    author = "Albakry, M. F. and others",
    collaboration = "SuperCDMS",
    title = "{Search for low-mass dark matter via bremsstrahlung radiation and the Migdal effect in SuperCDMS}",
    eprint = "2302.09115",
    archivePrefix = "arXiv",
    primaryClass = "hep-ex",
    reportNumber = "112013",
    doi = "10.1103/PhysRevD.107.112013",
    journal = "Phys. Rev. D",
    volume = "107",
    number = "11",
    pages = "112013",
    year = "2023"
}

@article{PICO_WIMP_2019,
    author = "Amole, C. and others",
    collaboration = "PICO",
    title = "{Dark Matter Search Results from the Complete Exposure of the PICO-60 C$_3$F$_8$ Bubble Chamber}",
    eprint = "1902.04031",
    archivePrefix = "arXiv",
    primaryClass = "astro-ph.CO",
    reportNumber = "FERMILAB-PUB-19-073-AE-E",
    doi = "10.1103/PhysRevD.100.022001",
    journal = "Phys. Rev. D",
    volume = "100",
    number = "2",
    pages = "022001",
    year = "2019"
}

@article{DarkSide_2023_PRD_SI,
    author = "Agnes, P. and others",
    collaboration = "DarkSide-50",
    title = "{Search for low-mass dark matter WIMPs with 12~ton-day exposure of DarkSide-50}",
    eprint = "2207.11966",
    archivePrefix = "arXiv",
    primaryClass = "hep-ex",
    reportNumber = "FERMILAB-PUB-22-589-ND-PPD-SCD",
    doi = "10.1103/PhysRevD.107.063001",
    journal = "Phys. Rev. D",
    volume = "107",
    number = "6",
    pages = "063001",
    year = "2023"
}

@article{EDELWEISS_WIMP_2019,
    author = "Armengaud, E. and others",
    collaboration = "EDELWEISS",
    title = "{Searching for low-mass dark matter particles with a massive Ge bolometer operated above-ground}",
    eprint = "1901.03588",
    archivePrefix = "arXiv",
    primaryClass = "astro-ph.GA",
    doi = "10.1103/PhysRevD.99.082003",
    journal = "Phys. Rev. D",
    volume = "99",
    number = "8",
    pages = "082003",
    year = "2019"
}

@article{XENONnT_2025_WIMP,
    author = "Aprile, E. and others",
    collaboration = "XENON",
    title = "{WIMP Dark Matter Search using a 3.1 tonne $\times$ year Exposure of the XENONnT Experiment}",
    eprint = "2502.18005",
    archivePrefix = "arXiv",
    primaryClass = "hep-ex",
    month = "2",
    year = "2025"
}

@article{CRESST_2019_WIMP,
    author = "Abdelhameed, A. H. and others",
    collaboration = "CRESST",
    title = "{First results from the CRESST-III low-mass dark matter program}",
    eprint = "1904.00498",
    archivePrefix = "arXiv",
    primaryClass = "astro-ph.CO",
    doi = "10.1103/PhysRevD.100.102002",
    journal = "Phys. Rev. D",
    volume = "100",
    number = "10",
    pages = "102002",
    year = "2019"
}

@article{LUKE1,
title = {Calorimetric ionization detector},
journal = {Nuclear Instruments and Methods in Physics Research Section A: Accelerators, Spectrometers, Detectors and Associated Equipment},
volume = {289},
number = {3},
pages = {406-409},
year = {1990},
issn = {0168-9002},
doi = {https://doi.org/10.1016/0168-9002(90)91510-I},
url = {https://www.sciencedirect.com/science/article/pii/016890029091510I},
author = {P.N. Luke and J. Beeman and F.S. Goulding and S.E. Labov and E.H. Silver}

}

@article{LUKE2,
  author  = {P. N. Luke},
  title   = {Voltage-assisted calorimetric ionization detector},
  journal = {Journal of Applied Physics},
  volume  = {64},
  number  = {12},
  pages   = {6858--6860},
  year    = {1988},
  month   = dec,
  doi     = {10.1063/1.341976},
  url     = {https://doi.org/10.1063/1.341976}
}

@article{energy_per_fission,
    author = "Zhang, Chao and Qian, Xin and Fallot, Muriel",
    title = "{Reactor antineutrino flux and anomaly}",
    eprint = "2310.13070",
    archivePrefix = "arXiv",
    primaryClass = "hep-ph",
    doi = "10.1016/j.ppnp.2024.104106",
    journal = "Prog. Part. Nucl. Phys.",
    volume = "136",
    pages = "104106",
    year = "2024"
}

@article{COHERENT:Ge_2025,
    author = "Adamski, S. and others",
    collaboration = "COHERENT",
    title = "{Evidence of Coherent Elastic Neutrino-Nucleus Scattering with COHERENT{\textquoteright}s Germanium Array}",
    doi = "10.1103/PhysRevLett.134.231801",
    journal = "Phys. Rev. Lett.",
    volume = "134",
    number = "23",
    pages = "231801",
    year = "2025"
}

@article{CONUS+_first_observation_2025,
    author = "Ackermann, N. and others",
    title = "{Direct observation of coherent elastic antineutrino{\textendash}nucleus scattering}",
    eprint = "2501.05206",
    archivePrefix = "arXiv",
    primaryClass = "hep-ex",
    doi = "10.1038/s41586-025-09322-2",
    journal = "Nature",
    volume = "643",
    number = "8074",
    pages = "1229--1233",
    year = "2025"
}

@article{geo_neutrino_SNOLAB_vs_CJPL,
  title={Experimental aspects of geoneutrino detection: Status and perspectives},
  author={Smirnov, Oleg},
  journal={Progress in Particle and Nuclear Physics},
  volume={109},
  pages={103712},
  year={2019},
  publisher={Elsevier}
}

@article{HFIR_fission_frac,
  title={Measurement of the Antineutrino Spectrum from U 235 Fission at HFIR with PROSPECT},
  author={Ashenfelter, J and Balantekin, AB and Band, HR and Bass, CD and Bergeron, DE and Berish, D and Bowden, NS and Brodsky, JP and Bryan, CD and Cherwinka, JJ and others},
  journal={Physical review letters},
  volume={122},
  number={25},
  pages={251801},
  year={2019},
  publisher={APS}
}

@article{migdal_neutrino_floor,
  title={A neutrino floor for the Migdal effect},
  author={Herrera, Gonzalo},
  journal={Journal of High Energy Physics},
  volume={2024},
  number={5},
  pages={1--17},
  year={2024},
  publisher={Springer}
}

@article{DayaBay_reactor_flux_uncertainty,
    author = "An, F. P. and others",
    collaboration = "Daya Bay",
    title = "{Evolution of the Reactor Antineutrino Flux and Spectrum at Daya Bay}",
    eprint = "1704.01082",
    archivePrefix = "arXiv",
    primaryClass = "hep-ex",
    doi = "10.1103/PhysRevLett.118.251801",
    journal = "Phys. Rev. Lett.",
    volume = "118",
    number = "25",
    pages = "251801",
    year = "2017"
}

\end{document}